\documentclass[fp,twocolumn]{jpsj3}
\usepackage{txfonts}
\usepackage{braket}
\usepackage{color}

\title{Possible Spin-Density Wave on Fermi Arc of Edge State \\ in Single-Component Molecular Conductors [Pt(dmdt)$_2$] and [Ni(dmdt)$_2$]}

\author{Taiki Kawamura$^{1}$, Biao Zhou$^{2}$, Akiko Kobayashi$^{2}$, and Akito Kobayashi$^{1}$}
\inst{$^{1}$Department of Physics, Nagoya University, Nagoya, Aichi 464-8602, Japan\\
$^{2}$Department of Chemistry, College of Humanities and Sciences, Nihon University, Setagaya-Ku, Tokyo 156-8550, Japan} 

\abst{Motivated by the highly one-dimensional edge state due to a Dirac nodal line the system in single-component molecular conductor [Pt(dmdt)$_2$], 
we investigate the magnetic properties of both [Pt(dmdt)$_2$] and [Ni(dmdt)$_2$], which are related materials by element substitution, 
by real-space-dependent random-phase approximation (RPA) in three-orbital Hubbard models with spin–orbit coupling, 
where these models are constructed using first-principles calculations.
We calculate longitudinal and transverse spin susceptibilities by three-dimensional real-space-dependent RPA.
We find that the helical spin-density wave (SDW) with incommensurate nesting of the Fermi arcs is induced at the edge by the Coulomb repulsion. 
We also find that the magnetic structure of the helical SDW can be changed by extremely small carrier doping, which is controllable
in molecular conductors.}

\begin{document}
\maketitle

\section{Introduction}
In condensed matter physics, Dirac electron systems are unconventional electronic states that have linear energy dispersions near the Fermi energy, which induce novel electronic properties such as quantum transport phenomena,\cite{T.Ando1998,V.P.Gusynin2006,S.Murakami2004,D.Hsieh2008} large diamagnetism,\cite{H.Fukuyama1970,H.Fukuyama2007} and anomalous magnetic responses.\cite{M.Hirata2016,M.Hirata2017,D.Ohki2020,M.HirataRPP} 
Graphene, for example, is an ideal two-dimensional massless Dirac electron system, where the $\pi$ electron on the graphene honeycomb sheet acts as a massless Dirac electron.\cite{P.R.Wallace1947,K.S.Novoselov2005}

Strongly correlated Dirac electron systems in two-dimensional organic conductors have been observed in $\alpha$-(BEDT-TTF)$_2$I$_3$ and similar substances via element substitution.\cite{K.Kajita1992,N.Tajima2000,A.Kobayashi2004,S.Katayama2006,A.Kobayashi2007,M.O.Goerbig2008,K.Kajita2014}
In fact, $\alpha$-(BEDT-TTF)$_2$I$_3$ exhibits a metal--insulator transition between the Dirac electron phase and the charge-ordered phase induced by nearest-neighbor Coulomb interactions,\cite{H.Seo2000,T.Takahashi2003,T.Takiuchi2007} where anomalous spin--charge separation on spin gaps \cite{K.Ishibashi2016,Y.Katayama2016} and transport phenomena occur.\cite{R.Beyer2016,D.Lu2016,D.Ohki2019} 
Furthermore, Coulomb interactions induce anomalous magnetic responses owing to the reshaping of the Dirac cone, ferrimagnetism, and spin-triplet excitonic fluctuations in the Dirac electron phase.\cite{M.Hirata2016,M.Hirata2017,D.Ohki2020,M.HirataRPP}
The effects of anions (I$_3^-$) and spin--orbit interactions (SOI) are also of interest to researchers.\cite{Alemany2012,S.M.Winter2017}

When the node of a two-dimensional Dirac electron system is extended in the three-dimensional direction, it becomes a Dirac nodal line. \cite{A.Burkov2011,C.-K.Chiu2014,C.Fang2015,Z.Gao2016} 
If the Dirac nodal line forms a closed curve, it is called a Dirac nodal ring system.
There are various Dirac nodal line/ring systems, such as graphite,\cite{P.R.Wallace1947} transition-metal monophosphates,\cite{H.Weng2015} Cu3N,\cite{Y.Kim2015} antiperovskites,\cite{R.Yu2015} perovskite iridates,\cite{J.-M.Carter2012} hexagonal pnictides CaAgX (X = P, As),\cite{A.Yamakage2016} and the single-component molecular conductors [Pd(dddt)$_2$] \cite{R.Kato2017_A,R.Kato2017_B,Y.Suzumura2017_A,Y.Suzumura2017_B,Y.Suzumura2018_A,Y.Suzumura2018_B,T.Tsumuraya2018,Y.Suzumura2019} and [Pt(dmdt)$_2$].\cite{B.Zhou2019,T.Kawamura2020}
Dirac nodal line/ring systems exhibit novel electronic properties that differ from those of two-dimensional systems, such as flat Landau levels,\cite{J.-W.Rhim2015} the Kondo effect,\cite{A.K.Mitchell2015} long-range Coulomb interaction,\cite{Y.Huh2016} and quasi-topological electromagnetic responses\cite{S.T.Ramamurthy2017}. 

Recently, the depletion of magnetic susceptibility has been observed in [Pt(dmdt)$_2$].\cite{B.Zhou2019} This anomalous magnetic response suggests the existence of electron correlation effects similar to those observed in the organic Dirac electron system $\alpha$-(BEDT-TTF)$_2$I$_3$.\cite{M.Hirata2016,M.Hirata2017,D.Ohki2020,M.HirataRPP}
Using a three-orbital tight-binding model based on density functional theory (DFT) calculations, we also discovered an extremely flat edge state that is topologically assigned on a specific surface of [Pt(dmdt)$_2$].\cite{T.Kawamura2020}
The local density of state (LDOS) at that specific edge has logarithmic peaks and very high values near the Fermi energy. This result strongly suggests magnetic instability at that specific edge. However, the magnetic properties have not yet been elucidated.

In this study, we investigate a possible edge-induced spin-density wave (SDW) in a three-orbital Hubbard model describing the Dirac nodal line systems of the single-component molecular conductors [Pt(dmdt)$_2$] and [Ni(dmdt)$_2$]. We observe incommensurate nesting on the Fermi arc of the edge state. Using real-space-dependent random-phase approximation (RPA), we show that, in the vicinity of charge neutral, SDW is induced on the expected surface because of nesting and Coulomb interactions. Because the strength of nesting and the nesting vector are quite sensitive to carrier doping, 
observations on the magnetic structure will precisely determine the amount of carrier doping.

In Sect. 2, we develop three-orbital tight-binding models of [Pt(dmdt)$_2$] and [Ni(dmdt)$_2$] based on the Wannier fitting of the first-principles calculation. In addition, we incorporate spin--orbit coupling (SOC) into the model based on the Kane--Mele model.\cite{Kane-Mele2005,T.Osada2018,T.Osada2019} We then revise the tight-binding model of [Pt(dmdt)$_2$] created in our previous study,\cite{T.Kawamura2020} because the revised model is more convenient than the old one. In Sect. 3, we derive the longitudinal and transverse spin susceptibilities on the basis of the real-space-dependent RPA of the Hubbard model. In Sect. 4, we calculate the edge state and LDOS of [Ni(dmdt)$_2$] at the edge. In Sect. 5, we calculate the longitudinal and transverse spin susceptibilities of [Pt(dmdt)$_2$] and [Ni(dmdt)$_2$] by real-space-dependent RPA, and we discuss the magnetic structure at the edge. In Sect. 6, a summary and a discussion are given. Finally, in Appendix, we discuss the validity of the three-orbital tight-binding model of [Ni(dmdt)$_2$].

\section{Tight-Binding Model}
In this section, we construct tight-binding models of the single-component molecular conductors [Pt(dmdt)$_2$] and [Ni(dmdt)$_2$] on the basis of the Wannier fitting of the first-principles calculation. The microscopic models of many molecular conductors such as the BEDT-TTF system can be constructed with one molecular orbit, which extends throughout a molecule. However, those of the single-component molecular conductors are constructed with multiple orbits in a molecule, which is called fragment orbits\cite{H.Seo2008,H.Seo2013,M.Tsuchiizu2011,M.Tsuchiizu2012}.
We have already developed a tight-binding model of [Pt(dmdt)$_2$] in our previous study\cite{T.Kawamura2020}. However, it is inconvenient to calculate the edge state and spin susceptibility. Therefore, we reconstruct that old model into a clear representation in this study. Because the model of [Ni(dmdt)$_2$] is formally the same as that of [Pt(dmdt)$_2$], we uniformly represent the tight-binding models of the two materials. Their differences are in hopping energies and site potentials. 

\begin{figure}[htpb]
\begin{center}
\includegraphics[width=60mm]{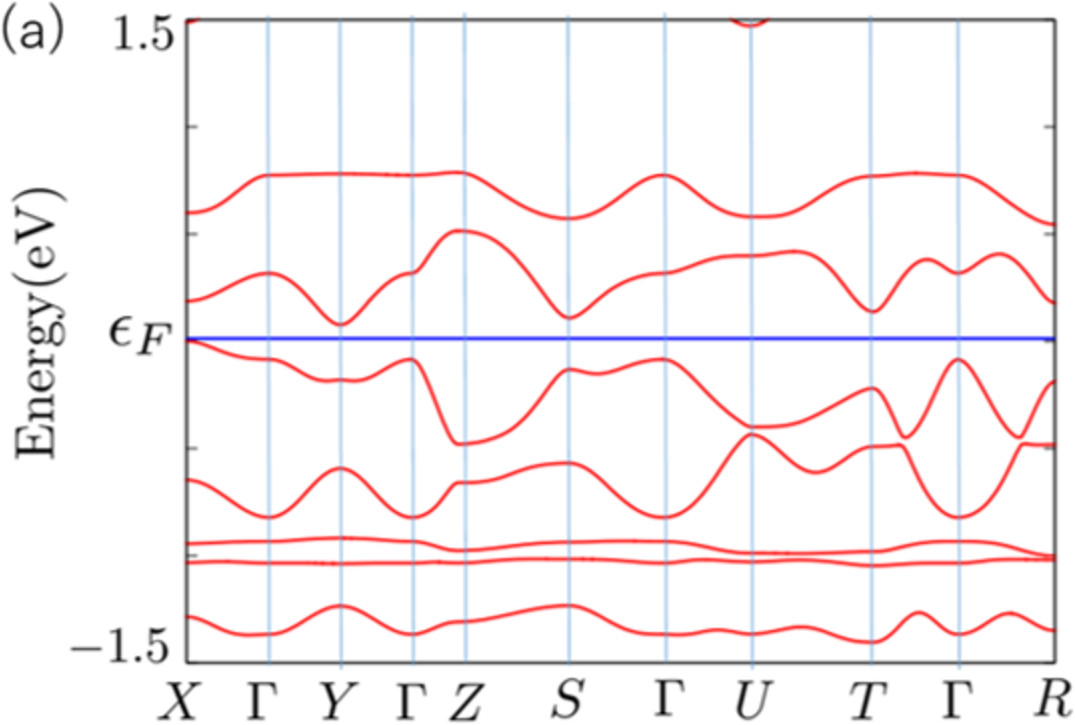}
\includegraphics[width=60mm]{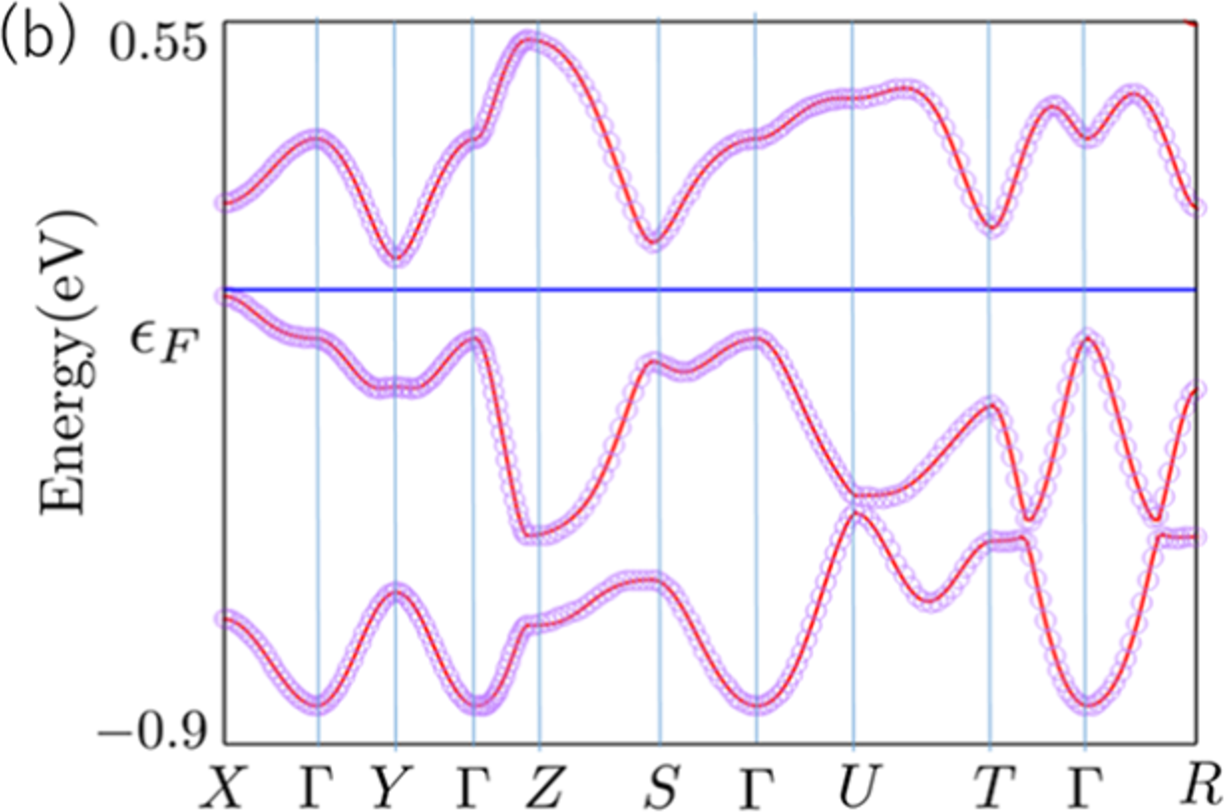}
\includegraphics[width=80mm]{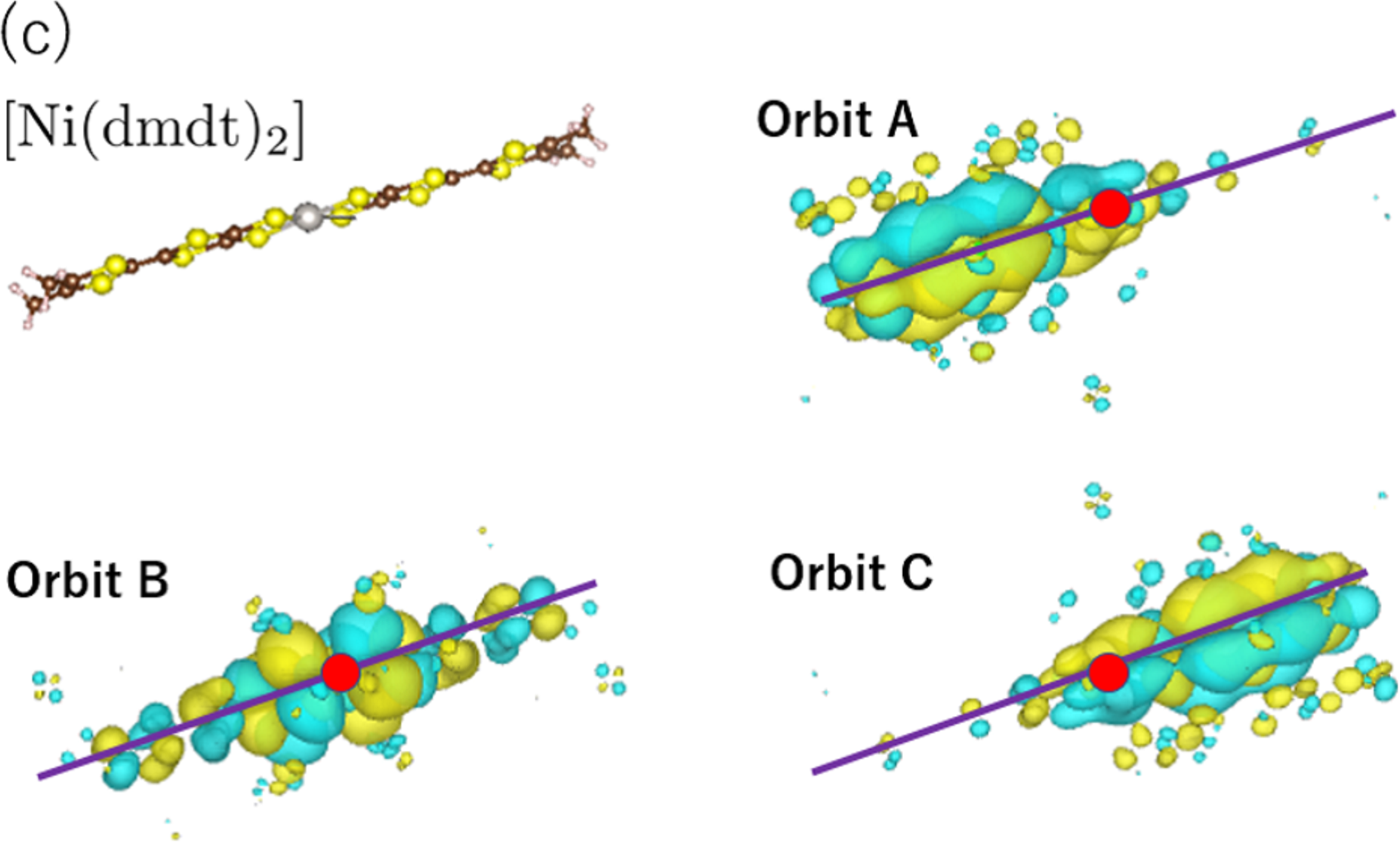}
\end{center}
\caption{(Color online) (a) Energy band structure of [Ni(dmdt)$_2$] determined by first-principles calculation. The horizontal axis represents connected symmetry points in BZ. 
(b) The red lines represent the band structure of [Ni(dmdt)$_2$] determined by first-principles calculation. The purple circles represent the energy band structure obtained by Wannier fitting. 
(c) [Ni(dmdt)$_2$] molecule and Wannier orbits A, B, and C. The red circle represents a Ni atom.}
\label{DFT}
\end{figure}

First, we calculate the energy band structure of [Ni(dmdt)$_2$] using Quantum ESPRESSO,\cite{P.Giannozzi2017} a first-principles calculation package, on the basis of the results of X-ray structure analysis. Figure~\ref{DFT}(a) shows the energy band structure determined by first-principles calculations. The horizontal axis connects the symmetry points in the first Brillouin zone (BZ), whereas the vertical axis is the energy measured from the Fermi energy. In our previous study,\cite{T.Kawamura2020} we determined that three isolated energy bands exist in [Pt(dmdt)$_2$]. However, because the corresponding three energy bands of [Ni(dmdt)$_2$] are close to the other bands, they can have degenerate points with the others. Therefore, we perform Wannier fitting on the six bands near the Fermi energy $\epsilon_{F}$ and construct a six-orbital tight-binding model using RESPACK,\cite{K.Nakamura2020} which is a software program used for Wannier fitting. However, we find that the three bands near the Fermi energy are not degenerate with the other bands and can be separated as in the model. We discuss more details of this phenomenon in Appendix. 

We perform Wannier fitting for the three bands and create the three-orbital tight-binding model again. Figure~\ref{DFT}(b) shows that Wannier fitting reproduces the first-principles calculation. The red lines and purple circles show the energy bands obtained by the first-principles calculation and Wannier fitting, respectively. Figure~\ref{DFT}(c) shows the Wannier orbits of [Ni(dmdt)$_2$], wherein orbits A(C) and B are distributed throughout the dmdt molecule and Ni atom, respectively. Orbit B is a \textit{d}-symmetry orbit, whereas orbits A and C are \textit{p}-symmetry orbits.

\begin{figure}[htpb]
\begin{center}
\includegraphics[width=70mm]{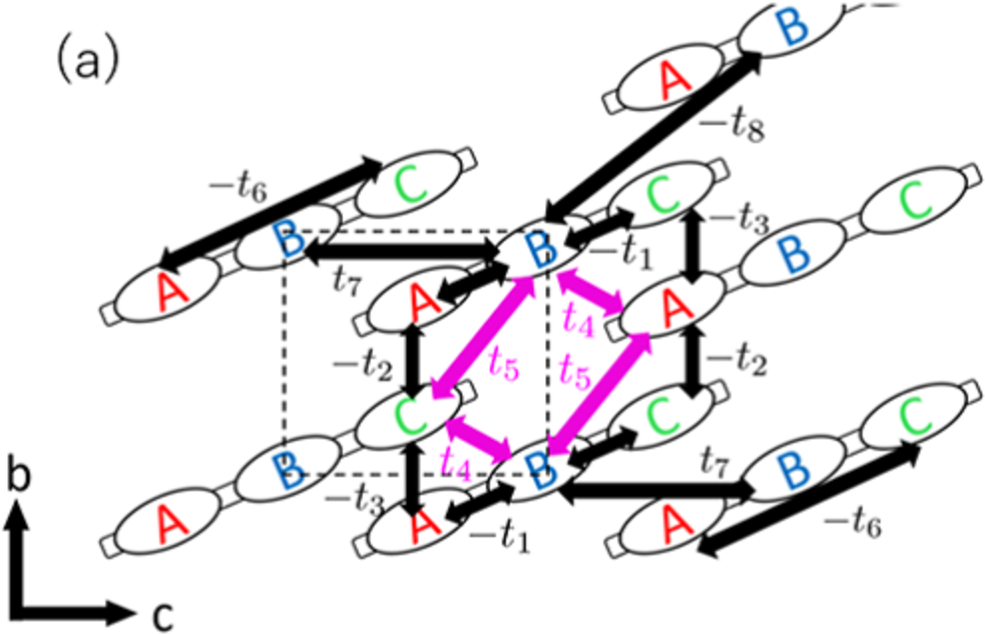}
\includegraphics[width=75mm]{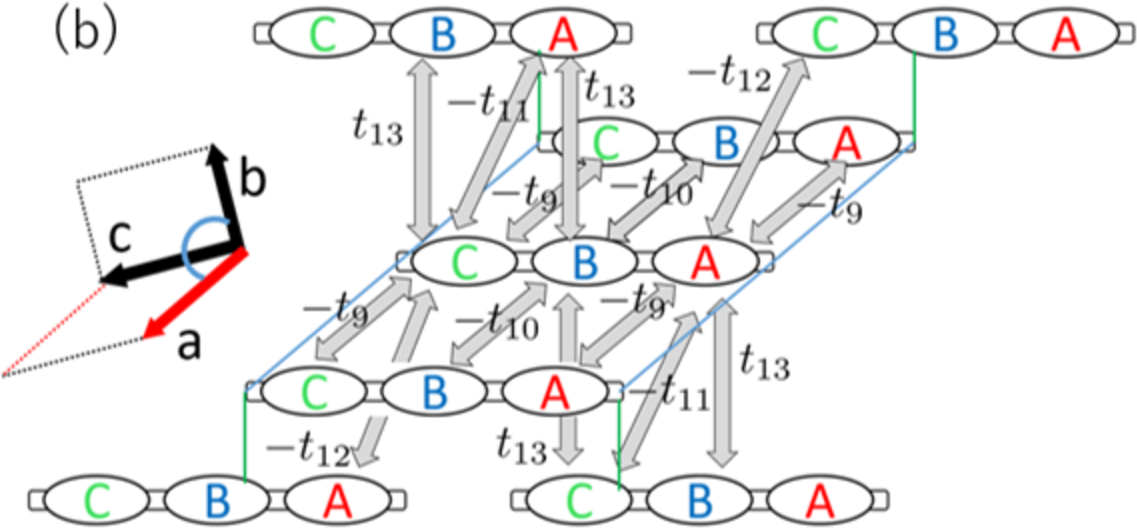}
\end{center}
\caption{(Color online)
(a) Schematic of the tight-binding model in the $b$-$c$ plane. The unit cell is shown by the broken line. (b) Schematic of the tight-binding model that includes hopping along the \textit{a}-axis. The blue and green lines are drawn to guide the eye. The blue lines are parallel to the $a$ direction. The green lines connect the molecules in the same $b$–$c$ plane.}
\label{net}
\end{figure}

The Hamiltonian of the tight-binding model incorporating SOC based on the Kane--Mele model\cite{Kane-Mele2005,T.Osada2018,T.Osada2019} is defined as
\begin{equation}
\begin{split}
H=&\sum_{<i,\alpha:j,\beta>,s}t_{i,\alpha;j,\beta}c^{\dagger}_{i,\alpha,s}c_{j,\beta,s}\\ 
&+i\lambda \sum_{<i,\alpha:j,\beta>,s}\nu_{i,\alpha:j,\beta}t_{i,\alpha;j,\beta}c^{\dagger}_{i,\alpha,s}\sigma^{z}_{ss}c_{j,\beta,s},
\end{split}
\end{equation}
where $s$ is the spin index, $i$ and $j$ represent unit cells, and $\alpha$ and $\beta$ are orbits. $t_{i\alpha;j\beta}$ is the hopping energy, which is obtained via Wannier fitting. 
$\lambda$ and $\sigma^{z}_{ss}$ are the SOC constant and diagonal component of the Pauli matrix $\hat{\sigma}^{z}$, respectively, for simplicity, because each unit cell has a flat molecule. $\nu_{i\alpha;j\beta}$ takes a value of either $0$ or $\pm 1$, which is determined by the outer product of the momenta and gradient of the potential energy $\textbf{P}\times\nabla U$. The qualitative band structures presented in this study are robust for any other treatment of SOC, although in a quantitative discussion, it will be necessary to treat the SOC constant and spin as vector quantities in consideration of the anisotropy of SOC\cite{Valenti}. By using the Fourier transform, we can rewrite Eq.~(1) as
\begin{equation}
H=\sum_{\textbf{k},\alpha,\beta,s}H(\textbf{k})_{\alpha,\beta,s}c^{\dagger}_{\textbf{k},\alpha,s}c_{\textbf{k},\beta,s},
\end{equation}
where \textbf{k} is the wavenumber vector. We obtain the hopping energies and site potentials via Wannier fitting of the first-principles calculation. Using the hopping energies, we create a three-orbital tight-binding model of [Ni(dmdt)$_2$]. In this study, we use hopping energies with absolute values that are larger than $0.010$ eV and incorporate SOC on the basis of the Kane--Mele model. Figure~\ref{net}(a) shows a schematic of the tight-binding model in the $b$-$c$ plane, whereas Fig.~\ref{net}(b) shows the hopping energies, which include hopping along the \textit{a}-axis. The magenta arrows(t$_4$ and t$_5$) represent the hopping energies, which are affected by SOC. In this study, the unit of energy is eV. With the hopping energies, the matrix elements of the Hamiltonian are expressed as
\begin{equation}
\begin{split}
H_{AA,s}(\textbf{k})&=-2t_{9}\cos{k_{a}}, \\ 
H_{AB,s}(\textbf{k})&= t_{13}e^{i(-k_{a}+k_{b}+k_{c})}-t_{1}\\ 
&+t_{5}(1+i\lambda\sigma)e^{i(k_{b}+k_{c})}+t_{4}(1-i\lambda\sigma)e^{ik_{c}},\\ 
H_{AC,s}(\textbf{k})&=-t_{11}e^{i(-k_{a}+k_{b}+k_{c})}-t_{12}e^{i(k_{a}+k_{c})}-t_{6}\\ 
&-t_{3}e^{ik_{c}}-t_{2}e^{i(k_{b}+k_{c})},\\ 
H_{BB,s}(\textbf{k})&= \Delta+2t_{7}\cos{(k_{b}+k_{c})}-2t_{8}\cos{k_{c}}\\
&-2t_{10}\cos{k_{a}},\\
H_{BC,s}(\textbf{k})&=t_{13}e^{i(-k_{a}+k_{b}+k_{c})}-t_{1}\\
&+t_{5}(1-i\lambda\sigma)e^{i(k_{b}+k_{c})}+t_{4}(1+i\lambda\sigma)e^{ik_{c}}, \\ 
H_{CC,s}(\textbf{k})&=-2t_{9}\cos{k_{a}},
\end{split}
\end{equation}
where $k_{a}$, $k_{b}$, and $k_{c}$ are the components of the wavenumber $\textbf{k}$. All lattice constants are defined to be 1. Because the Hamiltonian of [Ni(dmdt)$_2$] is formally the same as that of [Pt(dmdt)$_2$], we can calculate the electrical state of both materials with changes in only the hopping energies and site potential. Table I shows the hopping energies, which are larger than the cut-off energy of $0.01$ eV, and the site potentials of [Ni(dmdt)$_2$] and [Pt(dmdt)$_2$] obtained via Wannier fitting. The $t_{1}$, $t_{2}$, $\cdots$, and $t_{13}$ denote the hopping energies. We omit $t_{10}$ of [Ni(dmdt)$_2$] because it is less than the cut-off energy of $0.01$ eV. 
The $\Delta$ denotes the site potential of orbit B measured from those of orbits A and C.
\begin{table}[htb]
\begin{tabular}{|c|c|c|} \hline
hopping energy& [Ni(dmdt)$_2$] & [Pt(dmdt)$_2$] \\ \hline
$t_{1}$ & 0.237 & 0.212 \\
$t_{2}$ & 0.184 & 0.179 \\
$t_{3}$ & 0.208 & 0.201 \\
$t_{4}$ & 0.030 & 0.044 \\
$t_{5}$ & 0.033 & 0.043 \\
$t_{6}$ & 0.039 & 0.042 \\
$t_{7}$ & 0.010 & 0.014 \\
$t_{8}$ & 0.014 & 0.014 \\
$t_{9}$ & 0.014 & 0.024 \\
$t_{10}$ & -- & 0.013 \\
$t_{11}$ & 0.054 & 0.051 \\
$t_{12}$ & 0.053 & 0.051 \\
$t_{13}$ & 0.012 & 0.011 \\ \hline\hline
site potential & [Ni(dmdt)$_2$] & [Pt(dmdt)$_2$] \\ \hline
$\Delta$ & 0.043 & 0.070 \\ \hline
\end{tabular}
\caption{Hopping energies and site potentials of [Ni(dmdt)$_2$] and [Pt(dmdt)$_2$]. The $t_{10}$ of [Ni(dmdt)$_2$] is omitted because it is less than the cut-off energy $0.01$ eV.} 
\end{table}

In our previous study,\cite{T.Kawamura2020} we developed a tight-binding Hamiltonian of [Pt(dmdt)$_2$], which, however, is slightly difficult to use for the calculation of the edge state. On the other hand, Eq.~(3) is convenient for the calculation of the edge state because the orbits that belong to the same unit cell are in the same molecule.

\begin{figure}[htpb]
\begin{center}
\includegraphics[width=60mm]{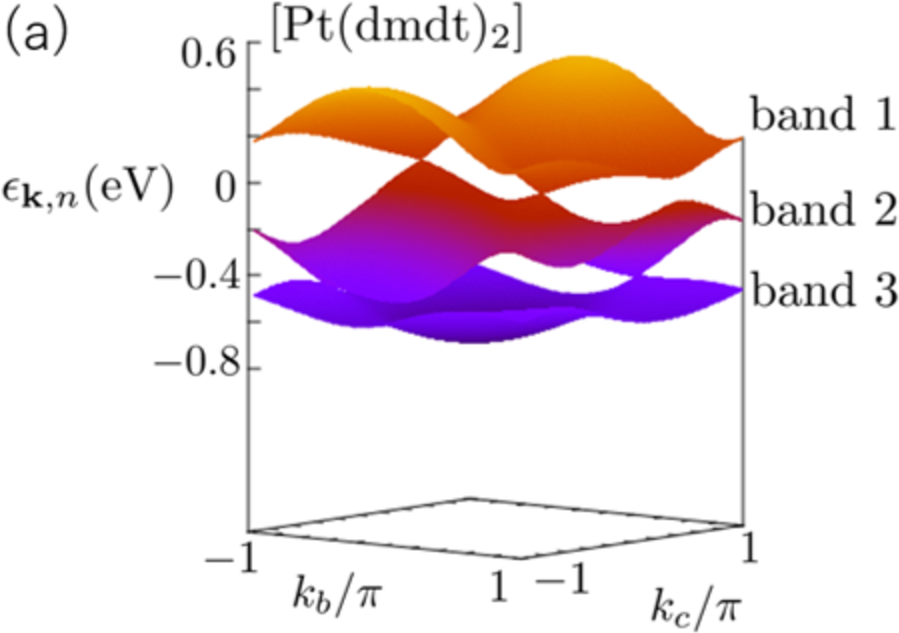}
\includegraphics[width=60mm]{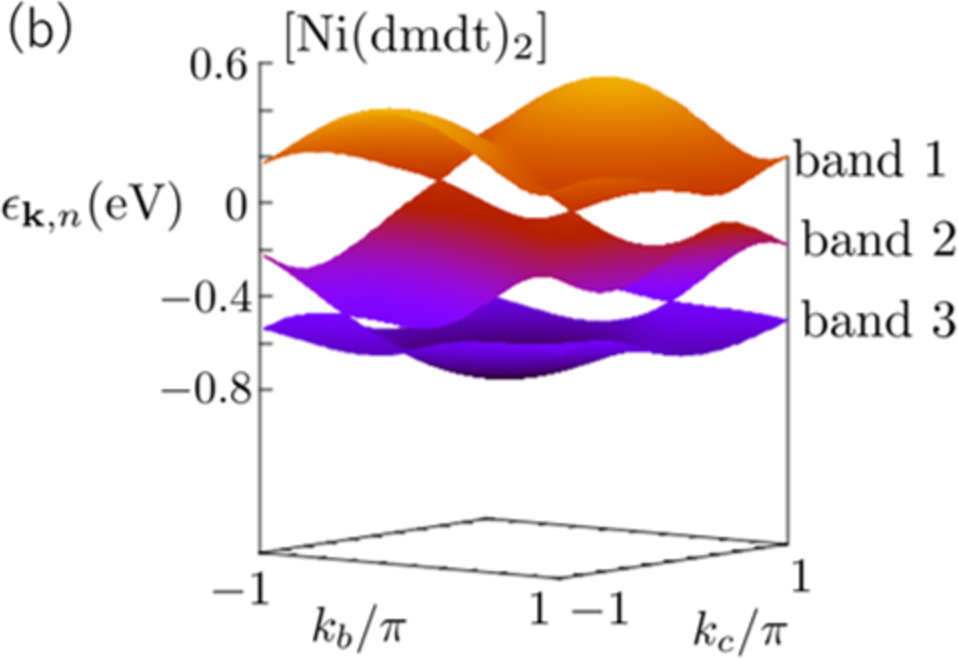}
\includegraphics[width=50mm]{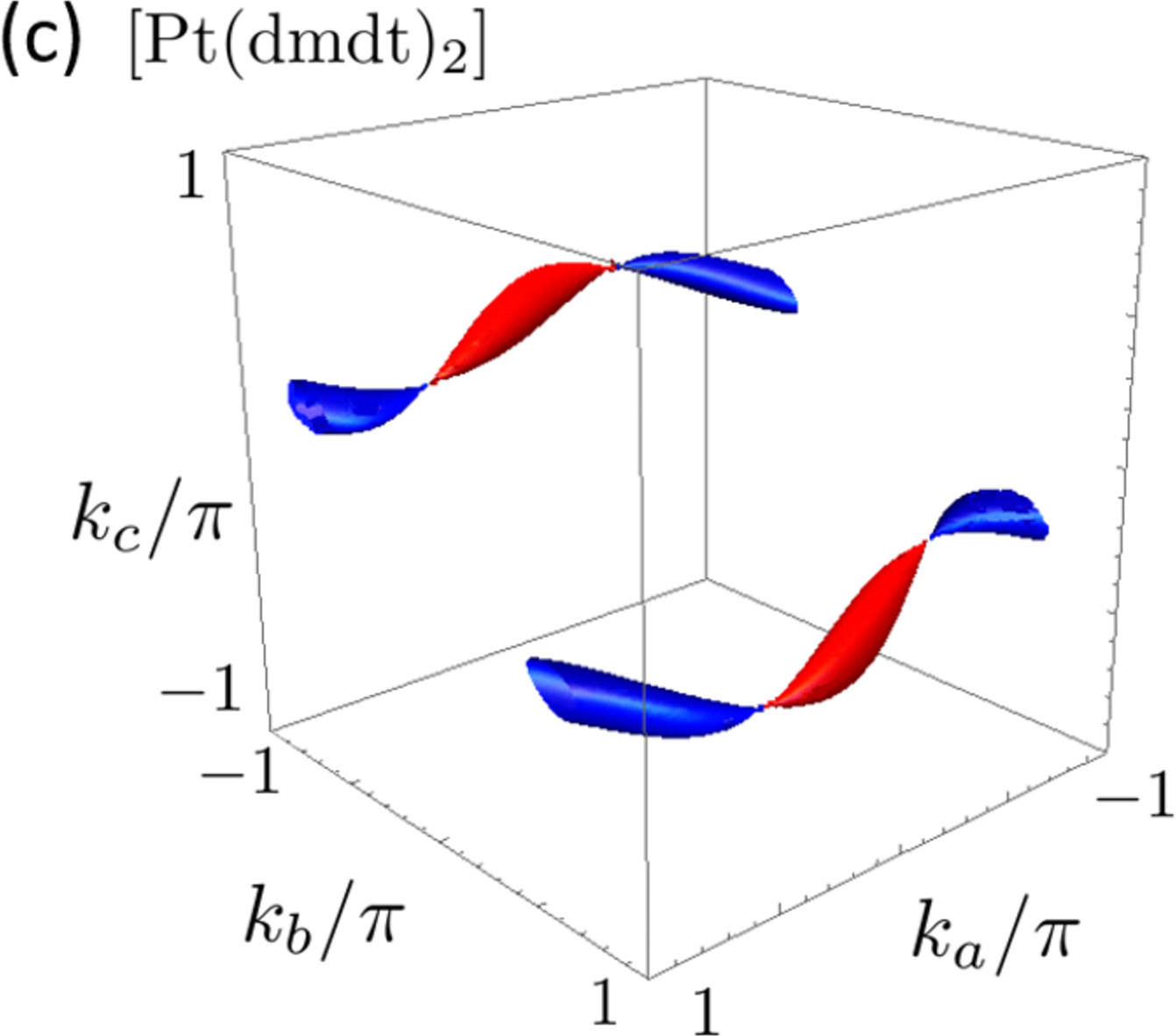}
\includegraphics[width=50mm]{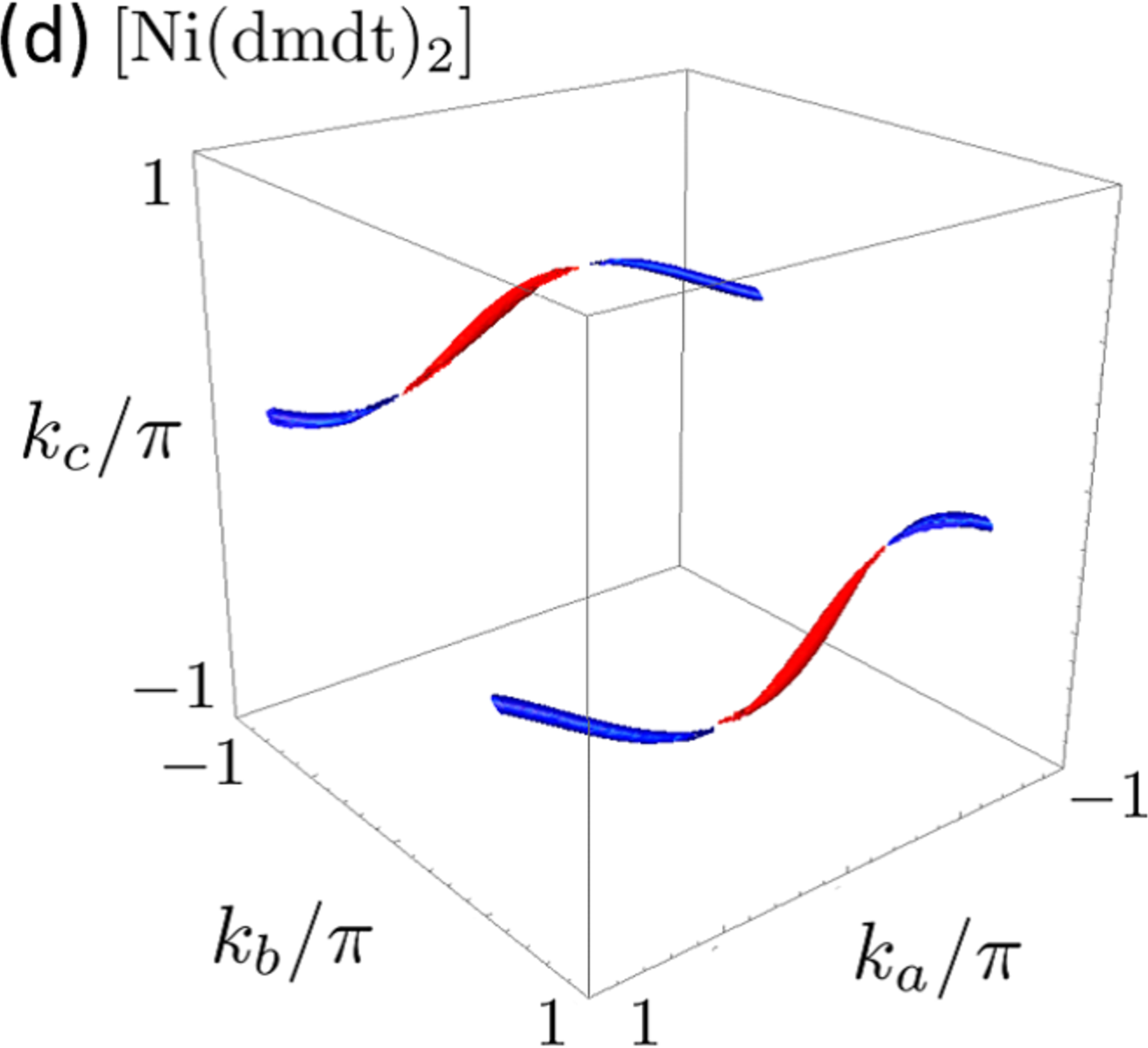}
\end{center}
\caption{(Color online)
(a) Energy dispersion of [Pt(dmdt)$_2$] in the $k_a$-$k_b$ plane calculated on the basis of the three orbits of the tight-binding model. The wavenumber is $k_a$=$-\pi/2$. 
(b) Energy dispersion of [Ni(dmdt)$_2$] in the $k_a$-$k_b$ plane calculated on the basis of the three orbits of the tight-binding model. The wavenumber is $k_a$=$-\pi/2$. 
(c) Fermi surface of [Pt(dmdt)$_2$] in the first BZ. 
(d) Fermi surface of [Ni(dmdt)$_2$] in the first BZ. 
The red Fermi surfaces inside the BZ and the blue ones on the boundary of the BZ are electron and hole pockets, respectively. 
}
\label{disp_FS}
\end{figure}

The Hamiltonian satisfies the energy eigenvalue equation
\begin{equation}
H_{s}(\textbf{k})\ket{{\textbf{k},n,s}}=E_{\textbf{k},n,s}\ket{{\textbf{k},n,s}},
\end{equation}
\[\ket{{\textbf{k},n,s}}=
\left(
\begin{array}{c}
d_{A,\textbf{k},n,s} \\
d_{B,\textbf{k},n,s} \\
d_{C,\textbf{k},n,s}
\end{array}
\right),
\]
where $\ket{{\textbf{k},n,s}}$ is the eigenvector, and $E_{\textbf{k},n,s}$ is the energy eigenvalue of the band $n$. On the othe hand,
$d_{\alpha,\textbf{k},n,s}$ is the wave function of site $\alpha$. 
In this study, $2/3$ of the energy band is filled. Therefore, the chemical potential $\mu$ is determined using 
\begin{equation}
\frac{1}{N_{L}}\sum_{\textbf{k},n,s} f_{\textbf{k},n,s}=4,
\end{equation}
where $f_{\textbf{k},n,s}$ is the Fermi distribution function, and $N_{L}$ is the number of unit cells. Here, we define the energy eigenvalue measured from the chemical potential $\epsilon_{\textbf{k},n}$:
\begin{equation}
\epsilon_{\textbf{k},n,s} \equiv E_{\textbf{k},n,s}-\mu.
\end{equation}
We obtain the energy dispersions of [Pt(dmdt)$_2$] and [Ni(dmdt)$_2$] by diagonalizing the Hamiltonian [Eq.~(3)] in the absence of SOC. Figures~\ref{disp_FS}(a) and \ref{disp_FS}(b) show the energy dispersions of [Pt(dmdt)$_2$] and [Ni(dmdt)$_2$], respectively, in the $k_b$-$k_c$ plane at $k_a=-\pi/2$. Dirac cones exist between the bands of both materials. The Dirac points between bands 1 and 2 form the Dirac nodal line, whereas those between bands 2 and 3 form the Dirac nodal ring. For this reason, it is difficult to construct a one- or two-orbital model. 

We then calculate the Fermi surfaces of [Pt(dmdt)$_2$] and [Ni(dmdt)$_2$]. Figures~\ref{disp_FS}(c) and \ref{disp_FS}(d) show the Fermi surfaces of [Pt(dmdt)$_2$] and [Ni(dmdt)$_2$], respectively, in the first BZ. The red and blue Fermi surfaces are hole and electron pockets, respectively, which result from Dirac points moving up and down across the Fermi energy as the wavenumber $k_a$ varies. The Fermi pockets of [Ni(dmdt)$_2$] are smaller than those of [Pt(dmdt)$_2$], because the hopping energies $t_9$ and $t_{10}$ of [Ni(dmdt)$_2$] are lower than those of [Pt(dmdt)$_2$]. 

\begin{figure}[htpb]
\begin{center}
\includegraphics[width=60mm]{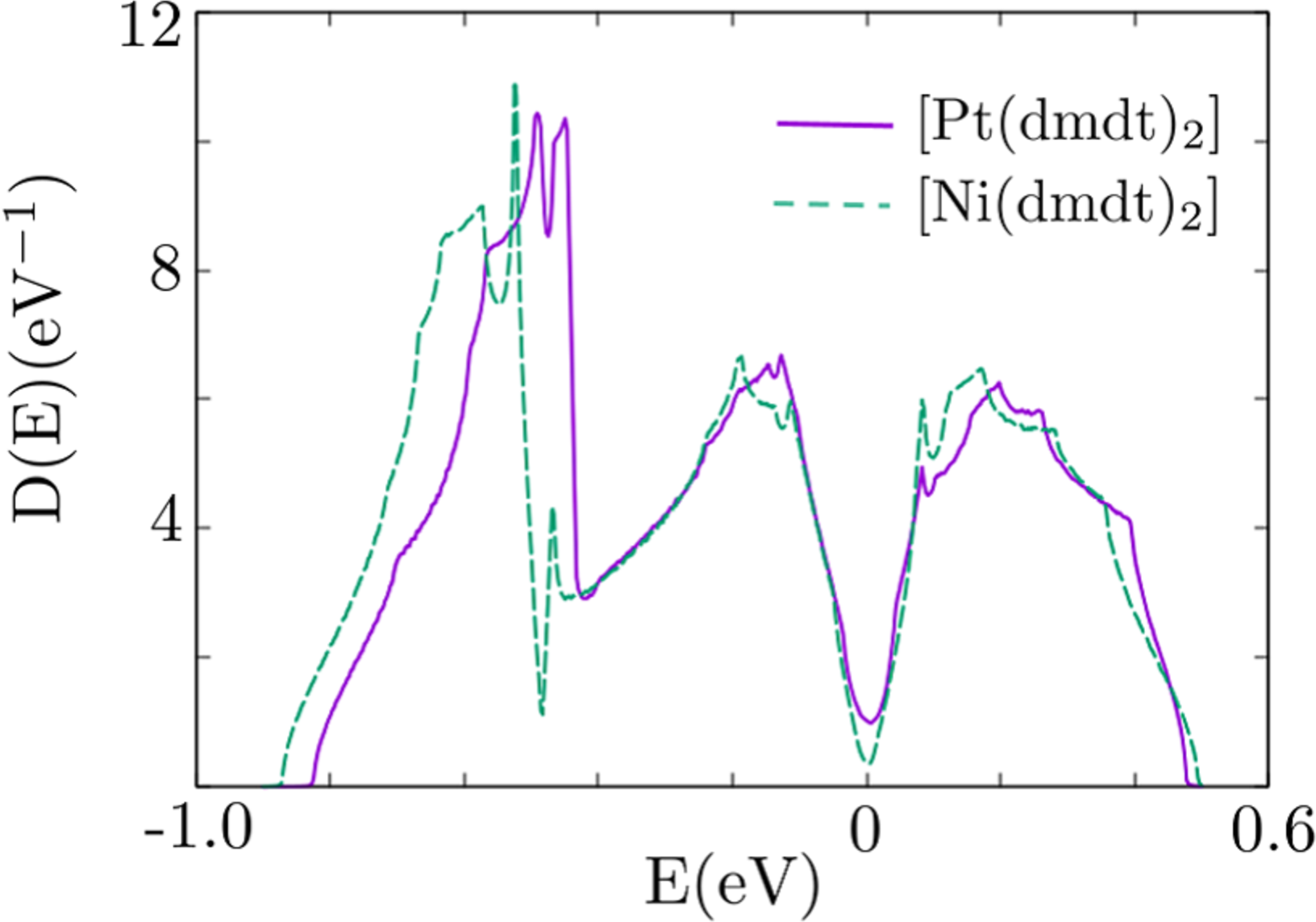}
\end{center}
\caption{(Color online) Purple and green lines represent DOSs of [Pt(dmdt)$_2$] and [Ni(dmdt)$_2$], respectively. The horizontal axis represents energy measured from the Fermi energy. 
}
\label{DOS}
\end{figure}

The density of state (DOS) is calculated to be 
\begin{equation}
D(E)=-\frac{1}{\pi N_L}\sum_{\textbf{k},n,s}\rm{Im}\left(\frac{1}{E-\epsilon_{\textbf{k},n,s}+i\eta}\right),
\end{equation}
where $\eta(>0)$ is an infinitesimally small value. Figure~\ref{DOS} shows the DOSs of [Pt(dmdt)$_2$] (purple line) and [Ni(dmdt)$_2$] (green line). The horizontal axis represents the energy measured from the Fermi energy. The DOSs have a valley near the Fermi energy because of the linear dispersion. In a two-dimensional Dirac electron system, $D(0)=0$. However, the DOSs of [Pt(dmdt)$_2$] and [Ni(dmdt)$_2$] are not zero at the Fermi energy because these materials have Fermi pockets. $D(E=0)$ reflects the size of the Fermi pockets, which suggests that [Ni(dmdt)$_2$] is more two-dimensional than [Pt(dmdt)$_2$]. 

The single-component molecular conductors [Pt(dmdt)$_2$] and [Ni(dmdt)$_2$] are regarded as the layer materials of the two-dimensional Dirac electron system such as graphene because linear dispersion is constructed in the $k_b$-$k_c$ plane and the hopping energies in the $a$ direction are smaller than  in the $k_b$-$k_c$ plane. The result that the Fermi pockets are small for the BZ suggests that [Pt(dmdt)$_2$] and [Ni(dmdt)$_2$] have high two-dimensionality. 
According to the first-principles calculation, it is considered that the wave functions of Pt in the [Pt(dmdt)$_2$] have little weight at the Fermi energy\cite{B.Zhou2019}. As a result, SOC of [Pt(dmdt)$_2$] is estimated to be smaller than the width of three bands by three orders magnitude. We assume that the SOC constant $\lambda$ of [Ni(dmdt)$_2$] is almost same as that of [Pt(dmdt)$_2$]. Thus, we consider that the Kane--Mele model is suitable for [Pt(dmdt)$_2$] and [Ni(dmdt)$_2$] in the present paper. In the realistic case, the SOC constants of [Pt(dmdt)$_2$] and [Ni(dmdt)$_2$] are estimated to be $\lambda$=$0.05$ ($\sim$$0.0022$ eV for [Pt(dmdt)$_2$] and $\sim$$0.0016$ eV for [Ni(dmdt)$_2$]).

According to RESPACK, the onsite Coulomb interactions of [Pt(dmdt)$_2$] and [Ni(dmdt)$_2$] are estimated to be about $5.4$ and $6.7$ eV, respectively. The onsite Coulomb interaction of [Ni(dmdt)$_2$] is slightly larger than that of [Pt(dmdt)$_2$]. They are much larger than the width of the three bands of both materials, which is about $1.3$ eV. Therefore, we consider that the electron correlation effect is important for both materials.

\section{Formulation}
The Hubbard model with SOC describing the electronic states of [Pt(dmdt)$_2$] and [Ni(dmdt)$_2$] is given by 
\begin{equation}
\begin{split}
H=&\sum_{<i,\alpha:j,\beta>,s}t_{i,\alpha;j,\beta}c^{\dagger}_{i,\alpha,s}c_{j,\beta,s}\\ 
&+i\lambda \sum_{<i,\alpha:j,\beta>,s,s'}\nu_{i,\alpha:j,\beta}t_{i,\alpha;j,\beta}c^{\dagger}_{i,\alpha,s}\sigma^{z}_{s,s'}c_{j,\beta,s'} \\
&+\sum_{i,\alpha}U_{\alpha}c^{\dagger}_{i,\alpha,\uparrow}c^{\dagger}_{i,\alpha,\downarrow}c_{i,\alpha,\downarrow}c_{i,\alpha,\uparrow}.
\end{split}
\end{equation}
In the case where the edge is cut for an axis, the Fourier transform cannot be implemented for that direction. By applying the Fourier transform on the other direction, we can rewrite Eq.~(8) as 
\begin{equation}
\begin{split}
&H=\sum_{\textbf{k},i\alpha,j\beta,s}H(\textbf{k})_{i\alpha,j\beta,s}c^{\dagger}_{\textbf{k},i\alpha,s}c_{\textbf{k},j\beta,s}\\
&+\frac{1}{N_L}\sum_{\textbf{k},\textbf{k}',\textbf{q},i\alpha}U_{i\alpha,i\alpha}c^{\dagger}_{\textbf{k}+\textbf{q},i\alpha,\uparrow}c^{\dagger}_{\textbf{k}'-\textbf{q},i\alpha,\downarrow}c_{\textbf{k}',i\alpha,\downarrow}c_{\textbf{k},i\alpha,\uparrow}\\
&\equiv H_{0}+H',
\end{split}
\end{equation}
where $i$ and $j$ are redefined as the numbers of unit cells along the direction that has the edge. The first term of Eq.~(9), which is $H_{0}$, is the tight-binding Hamiltonian that incorporates the SOC. $H(\textbf{k})_{i\alpha,j\beta,s}$ is obtained from Eq.~(3). The second term $H'$ represents the onsite Coulomb interaction, which acts between the up and down spins on the same site. $U_{i\alpha,i\alpha}$ is a constant value in the wavenumber space because it is the Fourier transform of the local Coulomb interaction in the real space. 

\begin{figure}[htpb]
\begin{center}
\includegraphics[width=90mm]{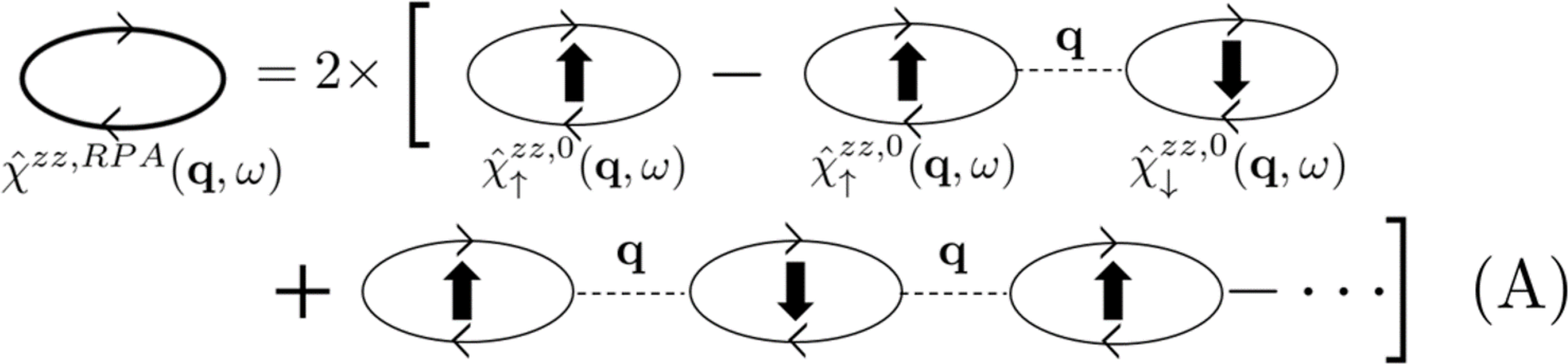}
\includegraphics[width=70mm]{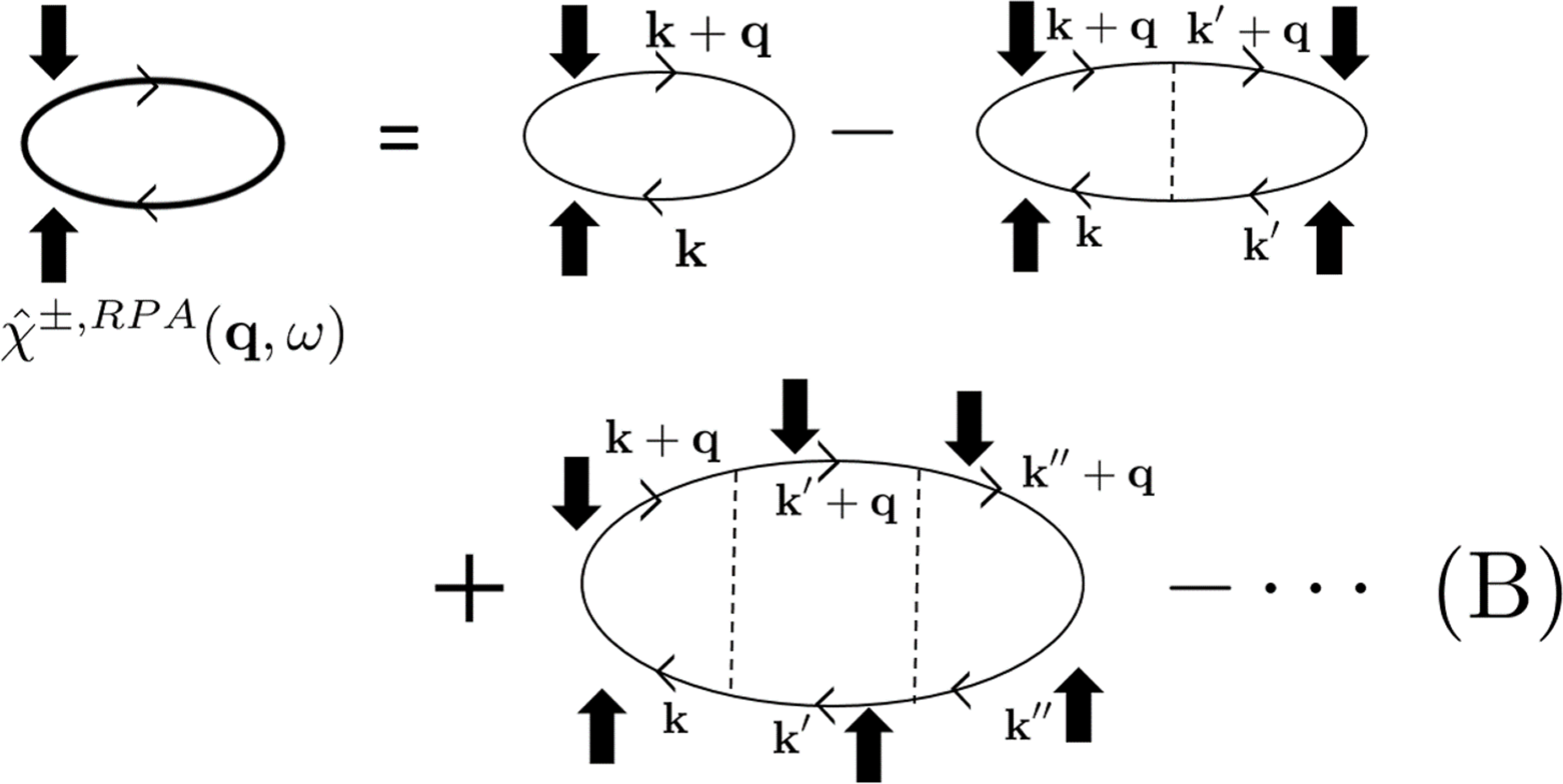}
\end{center}
\caption{(Color online)
Equation (A) is RPA for the longitudinal spin susceptibility $\chi^{zz,RPA}(\textbf{q},\omega)$. It is written on the basis of the ring Feynman diagram. 
Equation (B) is RPA for the transverse spin susceptibility $\chi^{\pm,RPA}(\textbf{q},\omega)$. It is written on the basis of the ladder Feynman diagram.}
\label{RPA}
\end{figure}

We then derive the longitudinal spin susceptibility by the three-dimensional real-space-dependent RPA\cite{Matsubara1,Matsubara2,Matsubara3}. According to the linear response theory, the longitudinal spin susceptibility matrix representing the real-space-dependence of longitudinal spin fluctuation is defined as
\begin{equation}
\hat{\chi}^{zz}(\textbf{q},i\omega_{m})=\int^{1/k_B T}_{0}d\tau e^{i\omega_m \tau}\left<T_{\tau}S^{z}_{\textbf{q}}(\tau)S^{z}_{-\textbf{q}}(0)\right>,
\end{equation}
\begin{equation}
S^{z}_{\textbf{q}}=\frac{1}{N_L}\sum_{\textbf{k}}\left(c^{\dagger}_{\textbf{k}+\textbf{q},\uparrow}c_{\textbf{k},\uparrow}-c^{\dagger}_{\textbf{k}+\textbf{q},\downarrow}c_{\textbf{k},\downarrow}\right),
\end{equation}
where $\omega_m$ is the Matsubara frequency, $k_B T$ is the energy of the temperature, and $\tau$ is the imaginary time. $S^{z}_{\textbf{q}}(\tau)$ is the operator that represents the magnetization and is written on the basis of the Heisenberg picture. When perturbation expansion is applied to Eq.~(10), the bare longitudinal spin susceptibility is obtained as
\begin{equation}
\begin{split}
&(\hat{\chi}^{zz,0}_{s}(\textbf{q},i\omega_{m}))_{i\alpha,j\beta} \\
&=-\frac{k_B T}{N_L}\sum_{\textbf{k},l}G^{0}_{i\alpha,j\beta,s}(\textbf{k}+\textbf{q},i\omega_{m+l})G^{0}_{j\beta,i\alpha,s}(\textbf{k},i\omega_{l}),
\end{split}
\end{equation}
\begin{equation}
\begin{split}
G^{0}_{i\alpha,j\beta,s}(\textbf{k},i\omega_{l})=\sum_{n}d^{*}_{i\alpha,n,s}(\textbf{k})d_{j\beta,n,s}(\textbf{k})\frac{1}{i\omega_{l}-\epsilon_{\textbf{k},n,s}},
\end{split}
\end{equation}
where $G^{0}_{i\alpha,j\beta,s}(\textbf{k},i\omega_{l})$ is the matrix element of the non-interacting Matsubara Green function, and $\omega_l$ is the Matsubara frequency. $\epsilon_{\textbf{k},n,s}$ is the energy eigenvalue of $H_{0}$, which is measured from the chemical potential, and $d_{j\beta,n,s}(\textbf{k})$ is the eigenvector of $H_0$. On the other hand, $s$ and $n$ are the spin and band indices, respectively. When Eqs.~(12) and (13) are organized and the analytic continuation is applied to the Matsubara frequency, the longitudinal bare spin susceptibility [Eq.~(10)] can be rewritten as 
\begin{equation}
\begin{split}
&(\hat{\chi}^{zz,0}_{s}(\textbf{q},\omega))_{i\alpha,j\beta}=-\frac{1}{N_L}\sum_{\textbf{k},n,m}\frac{f_{\textbf{k}+\textbf{q},n,s}-f_{\textbf{k},m,s}}{\epsilon_{\textbf{k}+\textbf{q},n,s}-\epsilon_{\textbf{k},m,s}-\hbar \omega-i\eta}\\
&\times d^{*}_{i\alpha,n,s}(\textbf{k}+\textbf{q})d_{j\beta,n,s}(\textbf{k}+\textbf{q})
d^{*}_{j\beta,m,s}(\textbf{k})d_{i\alpha,m,s}(\textbf{k})\\
&\equiv (\hat{\chi}^{zz,0}(\textbf{q},\omega))_{i\alpha,j\beta}.
\end{split}
\end{equation}
We define $\hat{\chi}^{zz,0}_{\uparrow}(\textbf{q},\omega)=\hat{\chi}^{zz,0}_{\downarrow}(\textbf{q},\omega)\equiv\hat{\chi}^{zz,0}(\textbf{q},\omega)$ in this study because the time reversal symmetry is protected. $m$ and $n$ are band indices, $f_{\textbf{k},m,s}$ is the Fermi distribution function, and $\eta=0^{+}$. When real-space-dependent RPA is applied to Eq.~(10), the longitudinal spin susceptibility is rewritten as Eq.~(A) in Fig.~\ref{RPA} on the basis of the Feynman diagram. The bubble diagram in Eq.~(A) represents the bare longitudinal spin susceptibility. The broken line in Eq.~(A) represents the interaction, which contains the factor $-1$ because of perturbation expansion. Therefore, on the basis of the real-space-dependent RPA, the longitudinal spin susceptibility is expressed as the summation of the infinite geometrical series of the bare longitudinal spin susceptibility $\hat{\chi}^{zz,0}(\textbf{q},\omega)$ and interaction $\hat{U}$. Thus, the longitudinal spin susceptibility is approximated to be
\begin{equation}
\begin{split}
\hat{\chi}^{zz,RPA}(\textbf{q},\omega)=2\hat{\chi}^{zz,0}(\textbf{q},\omega)[\hat{I}-\hat{U}\hat{\chi}^{zz,0}(\textbf{q},\omega)]^{-1}
\end{split},
\end{equation}
where $\hat{U}$ and $\hat{I}$ are the onsite Coulomb interaction matrix and unit matrix, respectively. The onsite Coulomb interaction $\hat{U}$ is the diagonal matrix. We consider the onsite Coulomb interactions $U_{i\alpha,i\alpha}$ for $\alpha$=$A,B,C$ as parameters, 
which are reflected as the ratio of the magnitude of interactions obtained by RESPACK.
For [Pt(dmdt)$_2$], we treat the diagonal
matrix elements of the onsite Coulomb interaction
as $U_{iA,iA}$=$U_{iB,iB}$=$U_{iC,i}$=$U$. 
Meanwhile, for [Ni(dmdt)$_2$], we
treat the case as $U_{iB,iB}$=$U$ and $U_{iA,iA}$=$U_{iC,iC}$=$0.79U$.
We note that the onsite Coulomb interactions obtained by RESPACK is too large 
to use in the RPA, because the RPA tends to overestimate the enhancement of spin fluctuations.
We then derive the transverse spin susceptibility by a real-space-dependent RPA. The transverse susceptibility $\hat{\chi}^{\pm}(\textbf{q},i\omega_{m})$ is defined as 
\begin{eqnarray}
&\hat{\chi}^{\pm}(\textbf{q},i\omega_{m})=\int^{1/k_B T}_{0}d\tau e^{i\omega_m \tau}\left<T_{\tau}S^{+}_{\textbf{q}}(\tau)S^{-}_{-\textbf{q}}(0)\right>,\\
&S^{+}_{\textbf{q}}=\frac{1}{N_L}\sum_{\textbf{k}}c^{\dagger}_{\textbf{k},\uparrow}c_{\textbf{k}+\textbf{q},\downarrow},\\
&S^{-}_{-\textbf{q}}=\frac{1}{N_L}\sum_{\textbf{k}}c^{\dagger}_{\textbf{k}+\textbf{q},\downarrow}c_{\textbf{k},\uparrow},
\end{eqnarray} 
where $S^{+}_{\textbf{q}}$ and $S^{-}_{-\textbf{q}}$ are the ladder operators of the spin angular momentum. $S^{+}_{\textbf{q}}(\tau)$ is written on the basis of the Heisenberg picture. The method for deriving the transverse spin susceptibility $\hat{\chi}^{\mp}(\textbf{q},i\omega_{m})$ involves changing only $S^{+}_{\textbf{q}}$ and $S^{-}_{-\textbf{q}}$ to $S^{-}_{\textbf{q}}$ and $S^{+}_{-\textbf{q}}$. 

In the same manner as with the bare longitudinal susceptibility, the bare transverse spin susceptibility is expressed as 
\begin{equation}
\begin{split}
&(\hat{\chi}^{\pm,0}(\textbf{q},\omega))_{i\alpha,j\beta}=-\frac{1}{N_L}\sum_{\textbf{k},n,m}\frac{f_{\textbf{k}+\textbf{q},n,\downarrow}-f_{\textbf{k},m,\uparrow}}{\epsilon_{\textbf{k}+\textbf{q},n,\downarrow}-\epsilon_{\textbf{k},m,\uparrow}-\hbar \omega-i\eta}\\
&\times d^{*}_{i\alpha,n,\downarrow}(\textbf{k}+\textbf{q})d_{j\beta,n,\downarrow}(\textbf{k}+\textbf{q})
d^{*}_{j\beta,m,\uparrow}(\textbf{k})d_{i\alpha,m,\uparrow}(\textbf{k}).
\end{split}
\end{equation} 
The method for deriving the bare transverse spin susceptibility $\hat{\chi}^{\mp,0}(\textbf{q},i\omega_{m})$ involves changing only the spin indices. By a real-space-dependent RPA, we can rewrite Eq.~(16) as Eq.~(B) in Fig.~\ref{RPA} on the basis of the Feynman diagram. Thus, the transverse spin susceptibility is approximated as
\begin{eqnarray}
&\hat{\chi}^{\pm,RPA}(\textbf{q},\omega)=\hat{\chi}^{\pm,0}(\textbf{q},\omega)[\hat{I}-\hat{U}\hat{\chi}^{\pm,0}(\textbf{q},\omega)]^{-1},\\
&\hat{\chi}^{\mp,RPA}(\textbf{q},\omega)=\hat{\chi}^{\mp,0}(\textbf{q},\omega)[\hat{I}-\hat{U}\hat{\chi}^{\mp,0}(\textbf{q},\omega)]^{-1}.
\end{eqnarray}
Hereafter, we discuss the case in which the frequency $\omega=0$. In the absence of SOC, $\chi^{zz}(\textbf{q})=\chi^{\pm}(\textbf{q})=\chi^{\mp}(\textbf{q})$ because of the SU(2) symmetry. However, in the presence of SOC, $\hat{\chi}^{zz}(\textbf{q})\ne\hat{\chi}^{\pm}(\textbf{q})\ne\hat{\chi}^{\mp}(\textbf{q})$. 
\section{Edge State}
\begin{figure}[htpb]
\begin{center}
\includegraphics[width=70mm]{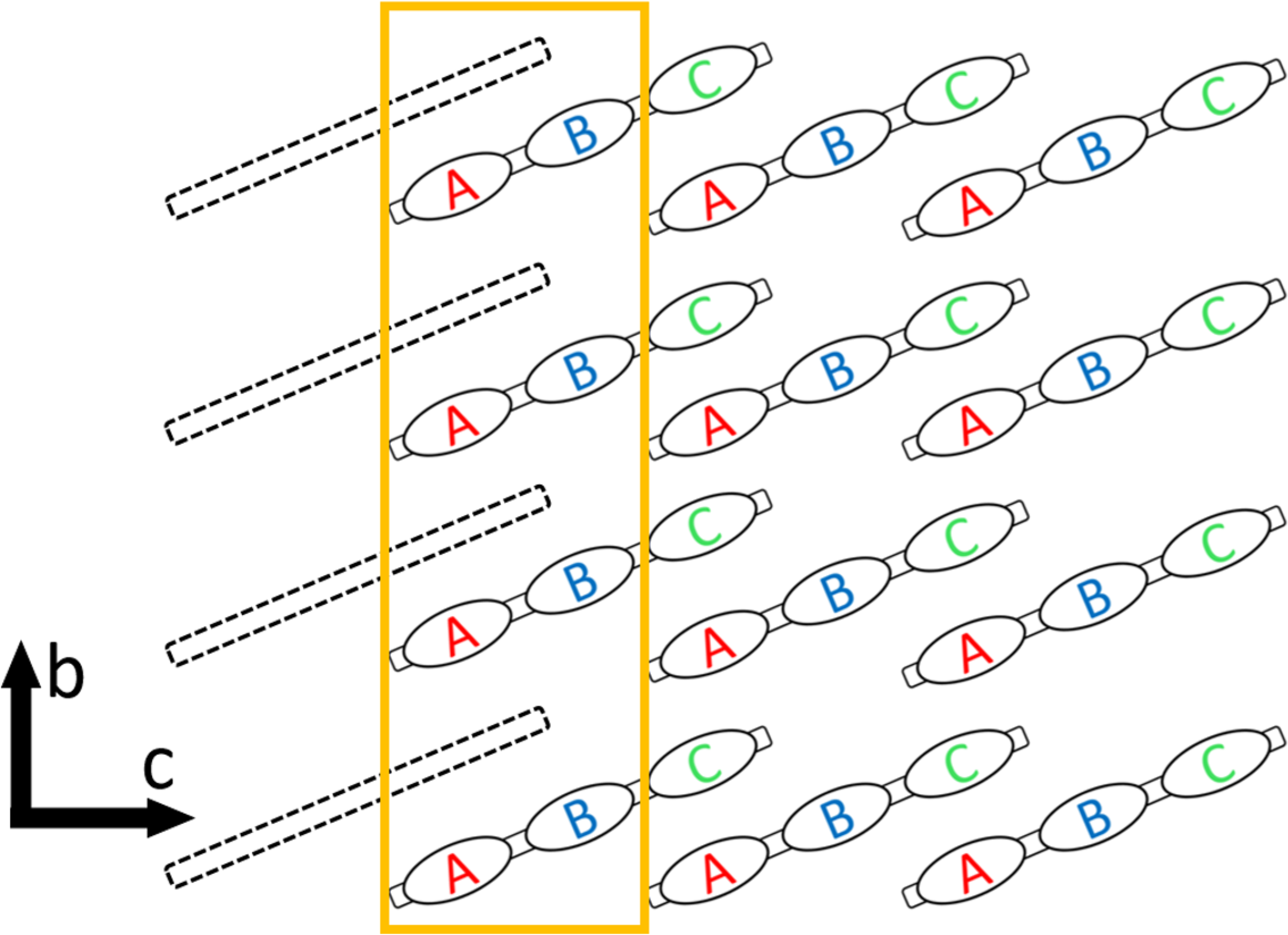}
\end{center}
\caption{(Color online) In this study, the edge is cut perpendicular to the \textit{c}-axis. The yellow box represents the orbits that act as the edge state at the $i$=$1$ edge. 
}
\label{cutting}
\end{figure}

\begin{figure}[htpb]
\begin{center}
\includegraphics[width=70mm]{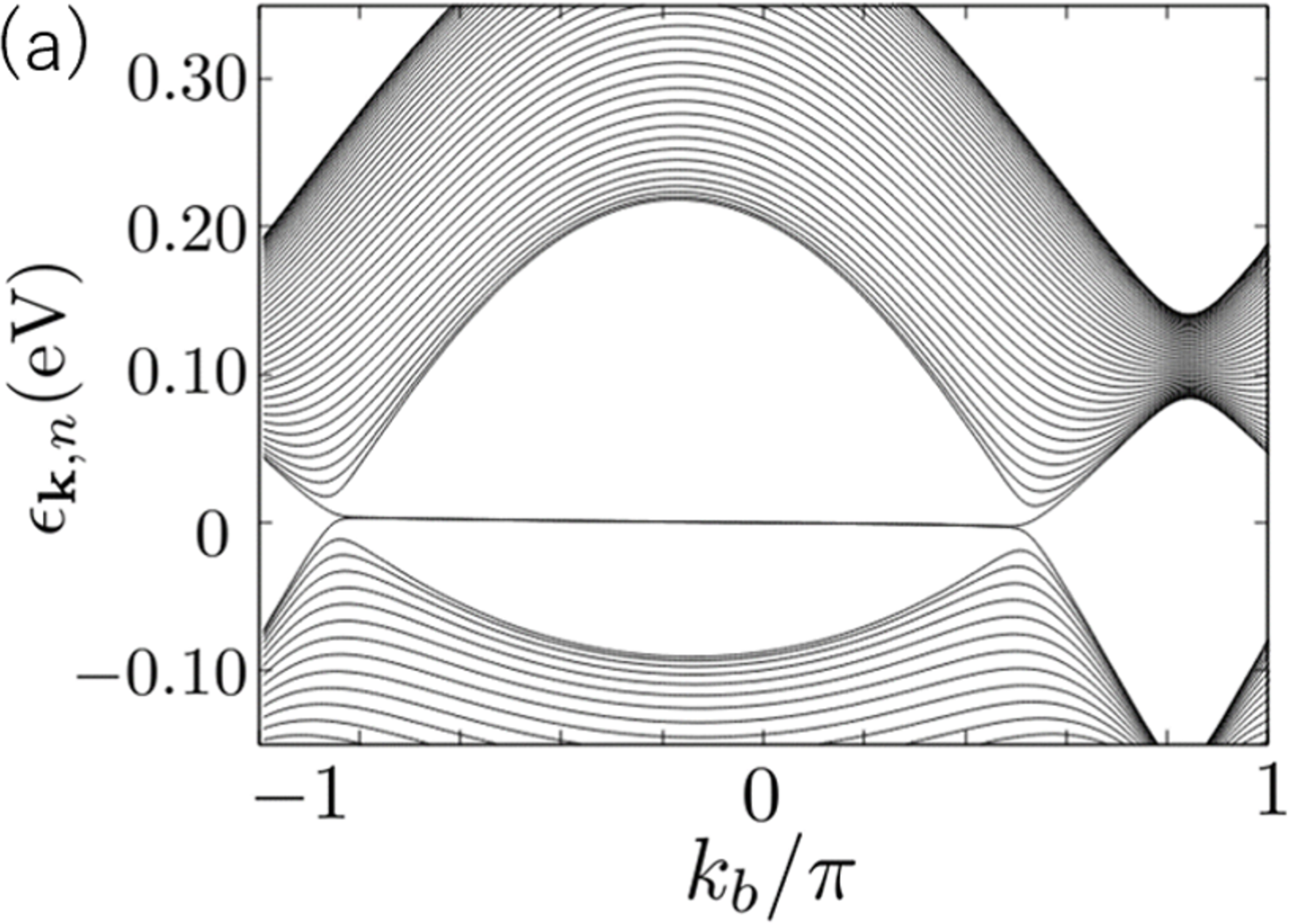}
\includegraphics[width=70mm]{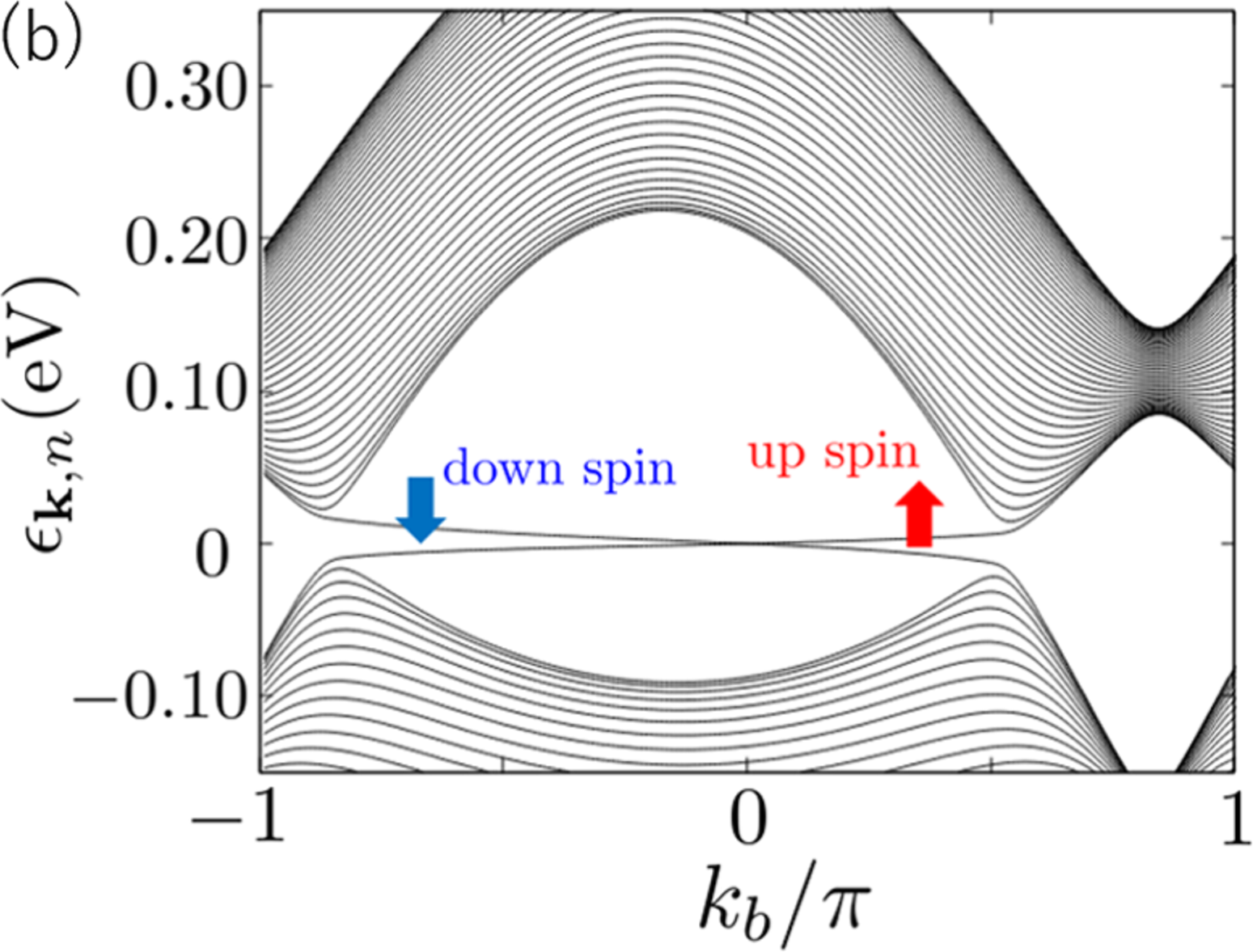}
\end{center}
\caption{(Color online)
(a) Energy dispersion of [Ni(dmdt)$_2$] at $k_a$=$-\pi/2$ in the cylindrical system (absence of SOC). The flat band corresponds to the edge state. 
(b) Energy dispersion of [Ni(dmdt)$_2$] at $k_a$=$-\pi/2$ in the cylindrical system (presence of SOC). SOC constant $\lambda$=$0.2$. The split bands correspond to the helical edge state. 
}
\label{disp_edge}
\end{figure}

In this section, we compare the topological nature of [Ni(dmdt)$_2$] with that of [Pt(dmdt)$_2$] in the non-interacting case. In our previous study\cite{T.Kawamura2020}, we found that the single-component molecular conductor [Pt(dmdt)$_2$] is a weak topological material that is
characterized by a topological number 0(100).
This indicates that [Pt(dmdt)$_2$] has an edge state, except at the $(100)$ edge. The parity eigenvalue is obtained when the spatial inversion matrix acts on the eigenvector of $H_{0}$ at the time-reversal invariant momenta (TRIMs). By using parity eigenvalues, we can calculate the topological number.\cite{FuKane,FuKaneMele,Moore,Roy,QiReview} Because the representation of the Hamiltonian in the present paper differs from that in our previous paper,\cite{T.Kawamura2020} the spatial inversion matrices in these two papers are also different. The spatial inversion matrix for the Hamiltonian in this study is 
\begin{eqnarray}
P (\textbf{k})= \left(
\begin{array}{ccc}
0 & 0 & 1 \\
0 & 1 & 0 \\
1 & 0 & 0
\end{array}
\right).
\end{eqnarray} 
With this spatial inversion matrix, the topological number $0(100)$ can be derived as in our previous paper.\cite{T.Kawamura2020} We calculate the energy dispersion of the system with the $(001)$ edge. A topological number of $0(100)$ indicates that the edge state occurs at the $(001)$ edge. Figure~\ref{cutting} shows a schematic of the $(001)$ edge. We cut the edge perpendicular to the \textit{c}-axis. The orbit is then represented by the expression $i\alpha$, where $i$ is the number of unit cells along the \textit{c}-axis, which is an integer $1\leq i \leq N_c$. On the other hand, $\alpha$ represents the Wannier orbits in a unit cell, which may be A, B, or C. By diagonalizing the Hamiltonian of the system with the (001) edge, we can obtain the energy dispersion shown in Fig.~\ref{disp_edge}(a). Specifically, Figs.~\ref{disp_edge}(a) and \ref{disp_edge}(b) show the energy dispersions ($k_a$=$-\pi/2$) of [Ni(dmdt)$_2$] in the absence and presence of SOC ($\lambda$=0.2$\sim$0.006 eV), respectively. The horizontal axis is for the wavenumber $k_b/\pi$, whereas the vertical axis is for the energy measured from the Fermi energy. In Fig.~\ref{disp_edge}(a), the flat band between the Dirac nodal lines is the energy dispersion of the edge state. Meanwhile, in Fig.~\ref{disp_edge}(b), the degenerate flat energy bands in Fig.~\ref{disp_edge}(a) split into two bands: one that has an up spin and the other has a down spin. This is because the energy gap is opened on the Dirac nodal line. The bands with up and down spins indicate the existence of a helical edge state. The combination of spins and bands depicted in Fig.~\ref{disp_edge}(b) is for the $i$=$1$ edge. The orbits that act as the edge state at edge $i$=$1$ are the orbits in the yellow frame in Fig.~\ref{cutting}. These are $i\alpha$=1A and 1B orbits. A similar result is obtained for [Pt(dmdt)$_2$]. 

\begin{figure}[htpb]
\begin{center}
\includegraphics[width=70mm]{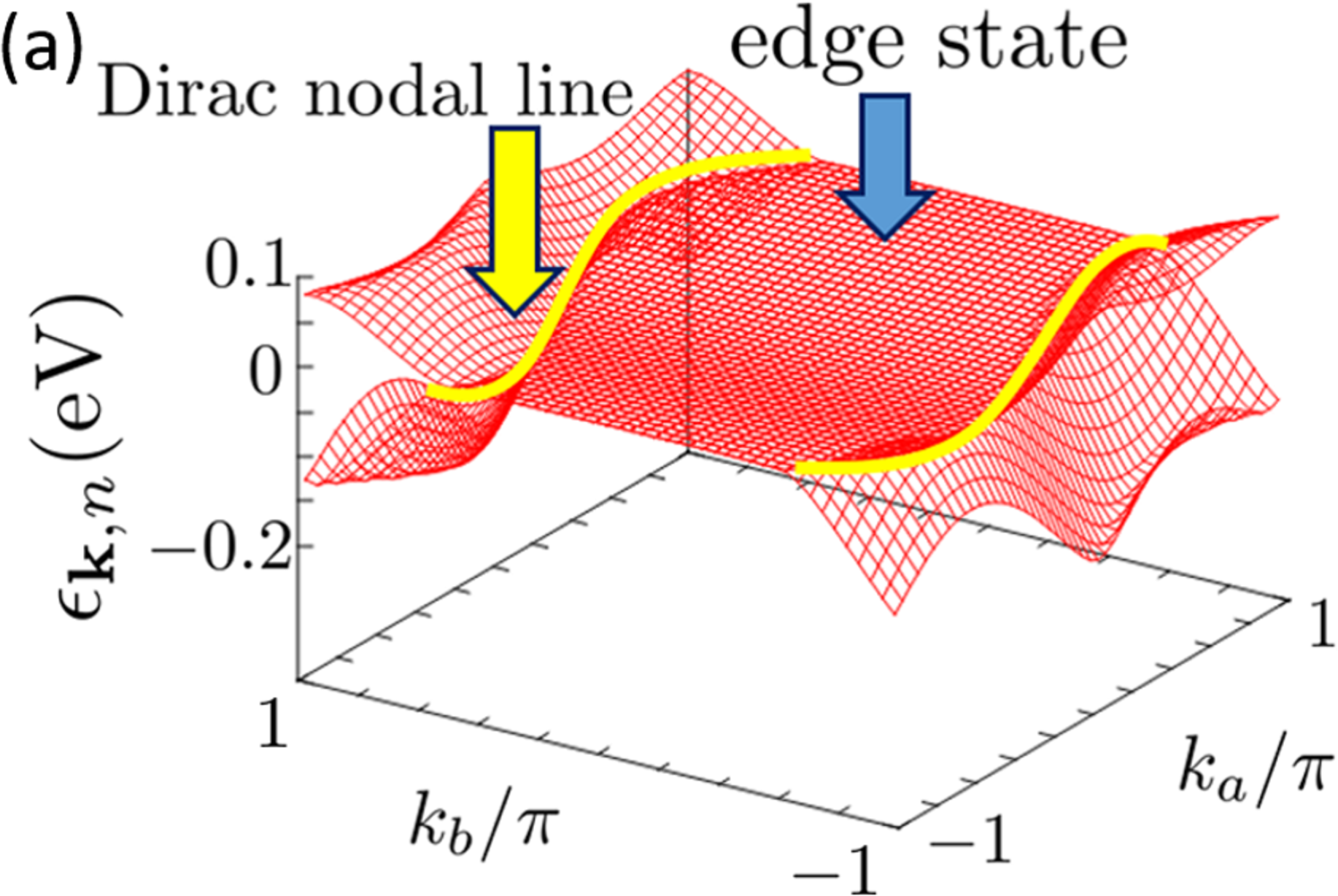}
\includegraphics[width=70mm]{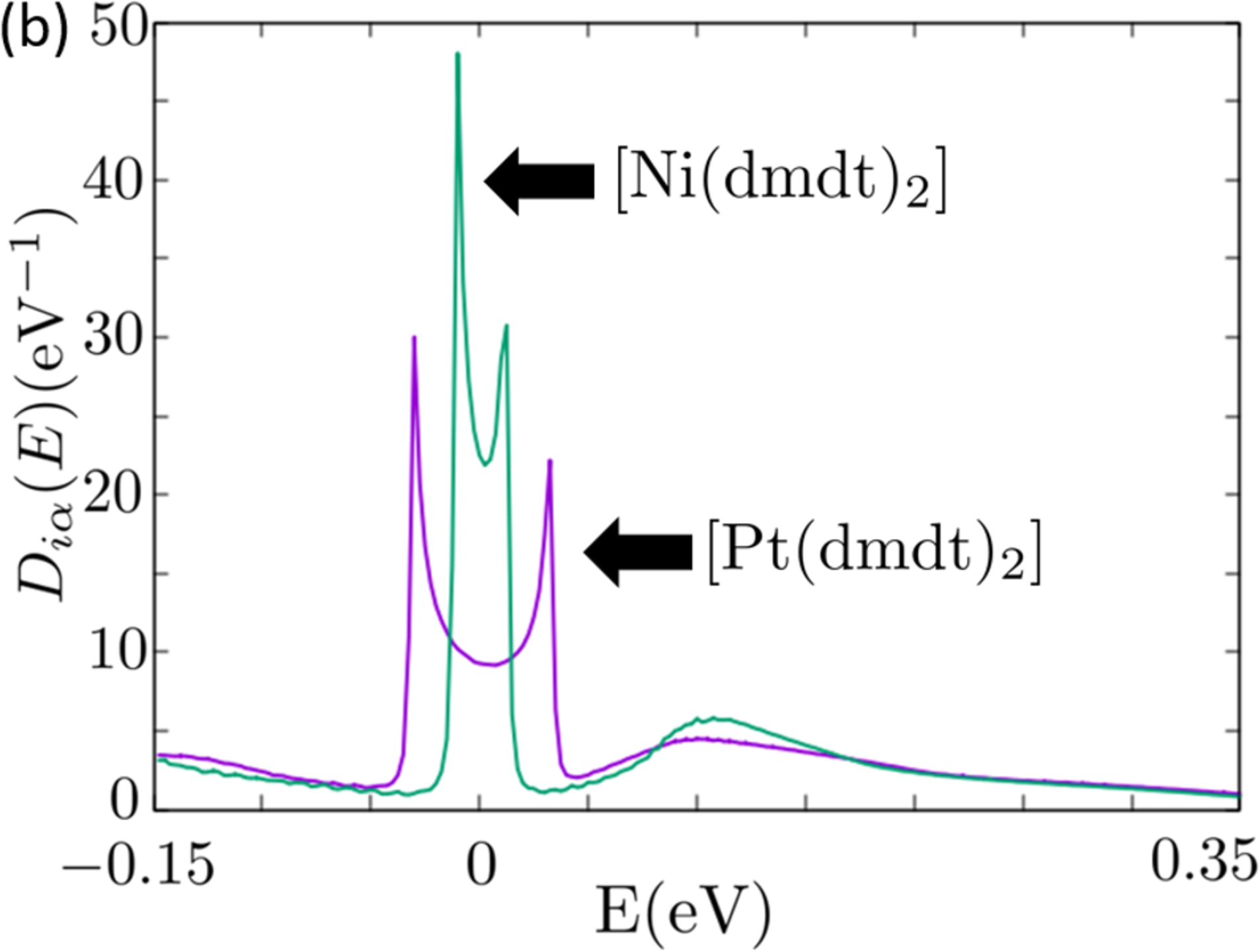}
\end{center}
\caption{(Color online)
(a) Energy dispersion of the edge state of [Ni(dmdt)$_2$] in $k_a$-$k_b$ plane (absence of SOC). The quasi-one-dimensional energy dispersion exists between the Dirac nodal lines, which corresponds to the energy dispersion of the edge state.
(b) The purple line represents the LDOS of [Pt(dmdt)$_2$]. The green line represents the LDOS of [Ni(dmdt)$_2$]. Logarithmic peaks occur near the Fermi energy.
}
\label{LDOS}
\end{figure}

Figure~\ref{LDOS}(a) shows the energy dispersion of the edge state of [Ni(dmdt)$_2$] in the $k_a$-$k_b$ plane in the absence of SOC. The edge state corresponds to the quasi-one-dimensional energy dispersion between the Dirac nodal lines. It results from the following two properties. The first property is the topological invariant. An analysis of the topological invariant reveals that the edge state appears along the \textit{a}-axis\cite{T.Kawamura2020}. The second property is the zigzag-like edge, which is perpendicular to the $c$-axis, and which we cut in this study. In graphene, the edge state at the zigzag edge is characterized by flat energy dispersion between the Dirac points,\cite{M.Fujita1996} suggesting that the edge states of [Pt(dmdt)$2$] and [Ni(dmdt)$2$] at the (001) edge correspond to the zigzag edge state of the graphene. 

We then calculate the local density of state (LDOS) at the edge. LDOS $D_{i\alpha}(E)$ is defined as
\begin{eqnarray}
&D_{i\alpha}(E)=\frac{1}{N_L}\sum_{\textbf{k}}A_{i\alpha}(\textbf{k},E),\\
&A_{i\alpha}(\textbf{k},E)=-\frac{1}{\pi}\sum_{s}{\rm Im}\hat{G}^0_{i\alpha,i\alpha,s}(\textbf{k},E+i\eta),
\end{eqnarray}
where $\hat{G}^0_{i\alpha,i\alpha,s}(\textbf{k},E+i\eta)$ is the retarded Green function, where the analytic continuation $i\omega_l \rightarrow E+i\eta$ is applied to Eq.~(13), and $\eta(>0)$ is an infinitesimally small value. $A_{i\alpha}(\textbf{k},E)$ is the spectral weight for the wavenumber $\textbf{k}$ and energy $E$. Figure~\ref{LDOS}(b) shows LDOSs $D_{1A}(E)+D_{1B}(E)$ at the edges of [Pt(dmdt)$_2$] and [Ni(dmdt)$_2$]. LDOSs have logarithmic peaks near the Fermi energy because of the quasi one dimensionality of the edge state. The logarithmic peaks of [Pt(dmdt)$_2$] and [Ni(dmdt)$_2$] correspond to approximately $\pm0.03$ and $\pm0.01$ eV, respectively, which correspond to the energies of the Fermi pockets of the respective materials. The large LDOS near the Fermi energy suggests that these materials have a magnetic structure at the (001) edge. The edge state at the (010) edge is also characterized by quasi-one-dimensional energy dispersion, but the LDOS is smaller at the (010) edge than that at the (001) edge. The edge state does not occur at the (100) edge, which is consistent with the topological number 0(100). Thus, in the next section, we discuss magnetism at the (001) edge induced by edge state and onsite Coulomb interaction. Long-range Coulomb interaction acts on the electrons in the graphene because Thomas--Fermi screening does not occur owing to the absence of DOS at the Fermi energy. This results in the reshaping of the Dirac cone, which is explained by the renormalization group theory.\cite{Gonz1994,Kotov} In particular, reshaping of the Dirac cone is realized in the organic conductor $\alpha$-(BEDT-TTF)$_2$I$_3$ and graphene.\cite{M.HirataReshaping} Moreover, the short-range Coulomb interaction results in the charge-ordered phase of the organic conductor $\alpha$-(BEDT-TTF)$_2$I$_3$ at a low pressure\cite{H.Seo2000,T.Takahashi2003,T.Takiuchi2007}. Thus, the effect of Coulomb interaction in organic conductors should be considered.

\section{Spin-Density Wave at Edge}
We calculate the three-dimensional real-space-dependent spin susceptibilities in the presence of the edge. Hereafter, we fix the temperature to $1.0$ K. Similar results are obtained at edges $i$=$1$ and $i$=$N_c$ because spatial inversion symmetry is protected. Therefore, we discuss the magnetism at the $i$=$1$ edge.
\subsection{Absence of spin--orbit coupling}
\begin{figure}[htpb]
\begin{center}
\includegraphics[width=70mm]{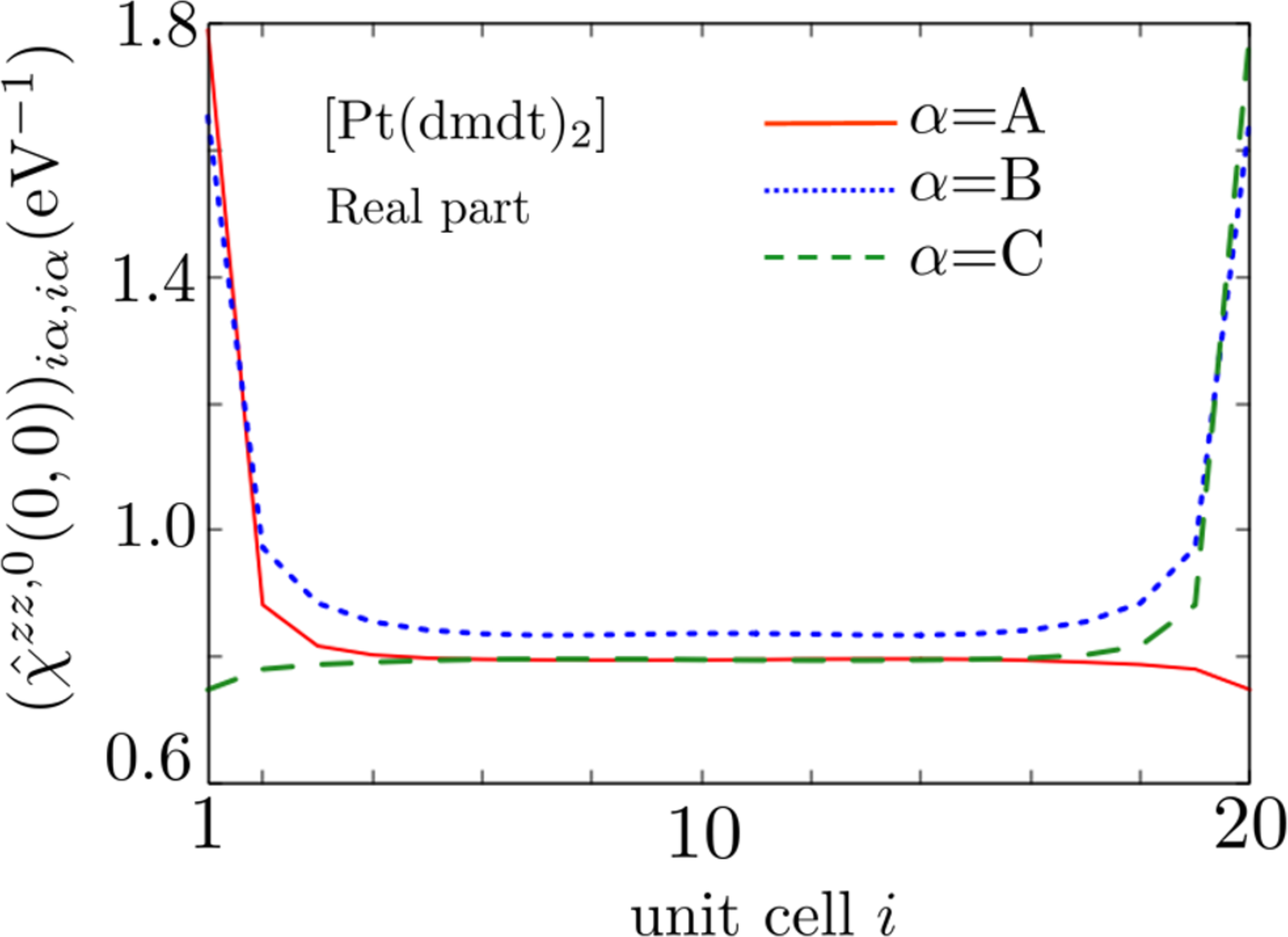}
\end{center}
\caption{(Color online) Unit cell $i$ dependence of $(\hat{\chi}^{zz,0}(\textbf{q}=\textbf{0}))_{i\alpha,i\alpha}$. The red, blue, and green lines represent $\alpha$=A, B, and C, respectively. At edges $i$=1 and $i$=$N_c$, we obtain equivalent results for commutation of orbits A and B because of the spatial inversion symmetry. Spin susceptibility at the edge is clearly higher than that of the bulk. We then set $N_c$=$20$.}
\label{chi0_uc}
\end{figure}

In the bulk, the DOSs of [Pt(dmdt)$_2$] and [Ni(dmdt)$_2$] are low near the Fermi energy because of linear energy dispersion (Fig.~\ref{DOS}). However, the LDOSs of the two materials at the (001) edge are high and have logarithmic peaks near the Fermi energy (Fig.~\ref{LDOS}). Thus, the edge magnetism can be enhanced. Using Eq.~(14), we calculate the unit cell $i$ dependence of the bare longitudinal spin susceptibility of $(\hat{\chi}^{zz,0}(\textbf{0}))_{i\alpha,i\alpha}$. Figure~\ref{chi0_uc} shows the unit cell $i$ dependence of $(\hat{\chi}^{zz,0}(\textbf{0}))_{i\alpha,i\alpha}$ of [Pt(dmdt)$_2$]. The susceptibilities of orbits A and B at the $i$=$1$ edge are larger than those in the bulk. The result for the $i$=$N_c$ edge is equivalent to that for the $i$=$1$ edge for the commutation of orbits A and C. Hence, we focus mainly on $i\alpha$=$1A, 1B$ in this study.

\begin{figure}[htpb]
\begin{center}
\includegraphics[width=75mm]{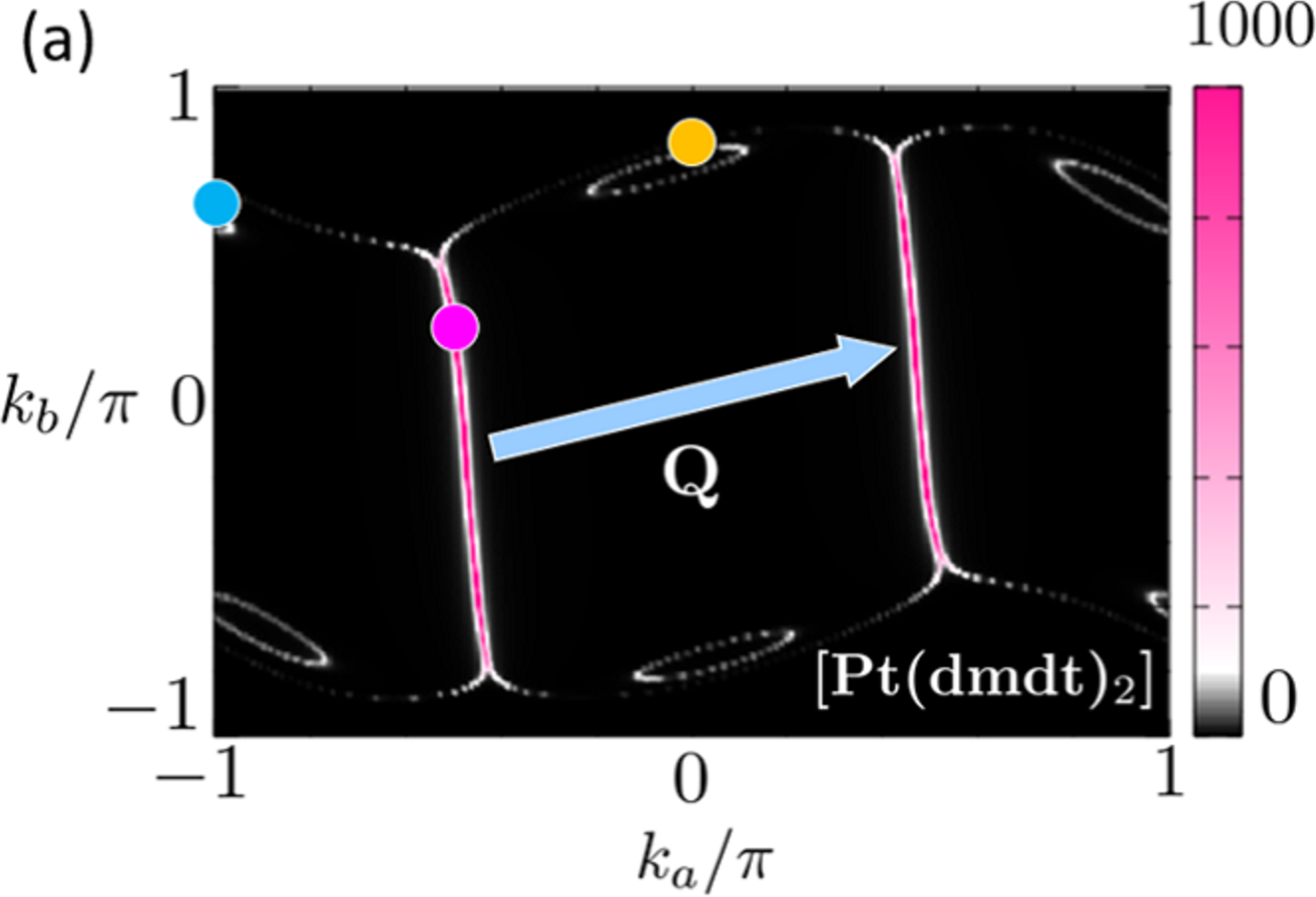}
\includegraphics[width=70mm]{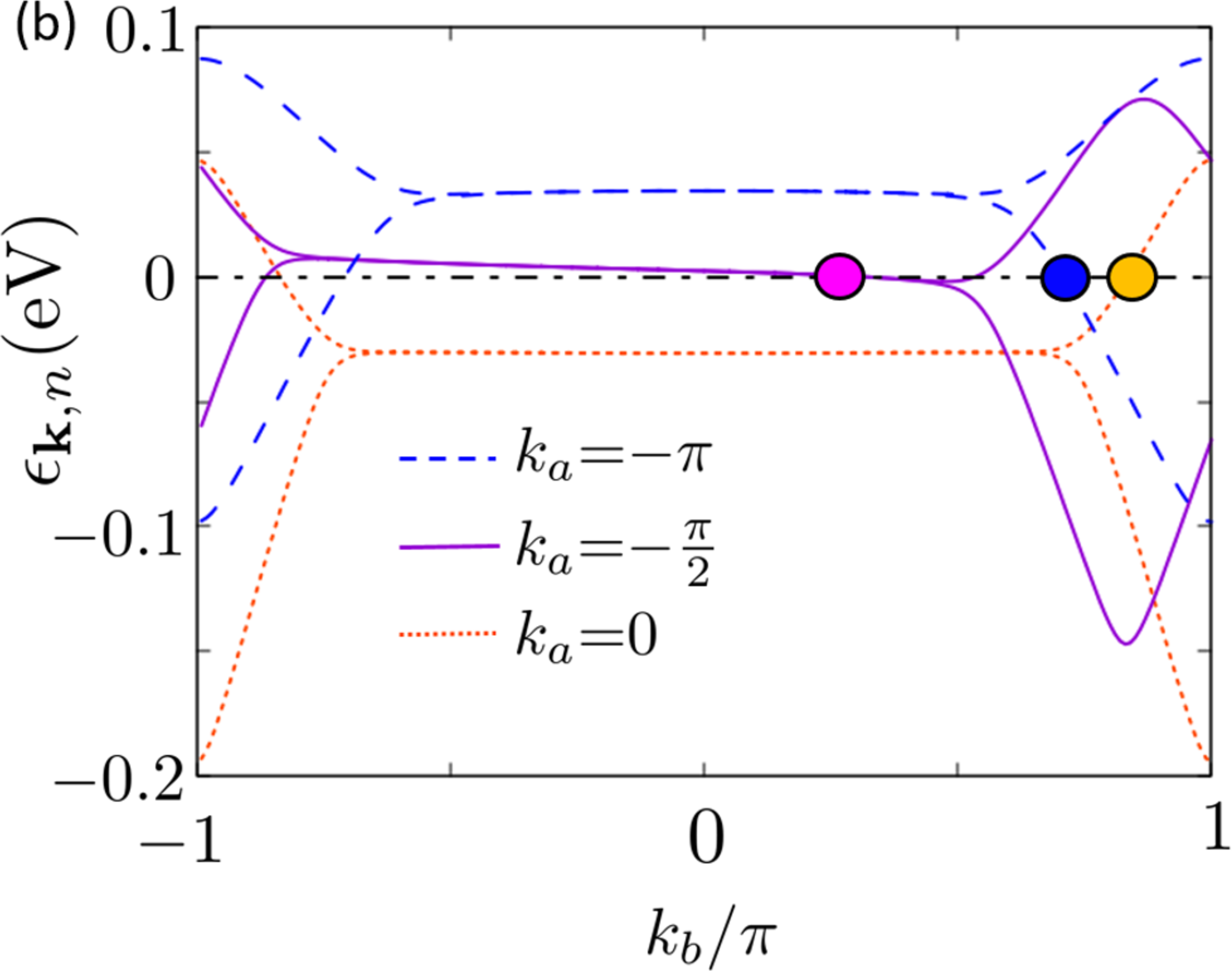}
\end{center}
\caption{(Color online) (a) Spectral weight for the Fermi energy of [Pt(dmdt)$_2$] at edge $i$=$1$, which is $A_{1A}(\textbf{k},0)+A_{1B}(\textbf{k},0)$. The two magenta lines correspond to the Fermi energy of the edge state. A good nesting vector \textbf{Q} connects the Fermi arc. (b) Energy dispersions of edge state of [Pt(dmdt)$_2$]. The flat bands correspond to the edge state. The blue, violet, and orange lines show the energy dispersions at $k_a$=$-\pi$, $k_a$=$-\pi/2$, and $k_a$=$0$, respectively. The magenta dot represents the intersection point of the edge state and Fermi energy, and it draws the magenta lines in (a). The blue and orange dots represent the intersection points of the bulk state and Fermi energy, respectively. 
}
\label{FS_nd}
\end{figure}

We then investigate the spin susceptibility at the (001) edge in detail. We calculate the Fermi arc at the edge. Figure~\ref{FS_nd}(a) shows the spectral weight for the Fermi energy of [Pt(dmdt)$_2$] at the edge. It is calculated using $A_{1A}(\textbf{k},0)+A_{1B}(\textbf{k},0)$ in Eq.~(24). The magenta straight lines represent the Fermi arc of the edge state of [Pt(dmdt)$_2$]. They have a good nesting vector \textbf{Q}. Meanwhile, the white line is the Fermi arc due to the energy dispersion of the bulk, which does not play an important role in this study. The spectral weight of the edge state is large because the wave function of the edge state is localized at the edge. By contrast, the spectral weight of the bulk state is small because the Bloch wave is not localized. It is predicted that the longitudinal susceptibility is enhanced by the nesting vector \textbf{Q}. In [Ni(dmdt)$_2$], an equivalent result is obtained. Figure~\ref{FS_nd}(b) shows the energy dispersions of the edge state of [Pt(dmdt)$_2$], where the blue, violet, and orange lines represent $k_a$=$-\pi$, $k_a$=$-\pi/2$, and $k_a$=$0$, respectively, and the horizontal axis represents $k_b$. The dots in Fig.~\ref{FS_nd} depict the intersection of the energy dispersion and the Fermi energy. These points correspond to the dots in Fig.~\ref{FS_nd}(a). The magenta dot is a part of the Fermi arc of the edge state, whereas the blue and orange dots are parts of the Fermi arc of the bulk state. 

\begin{figure}[htpb]
\begin{center}
\includegraphics[width=70mm]{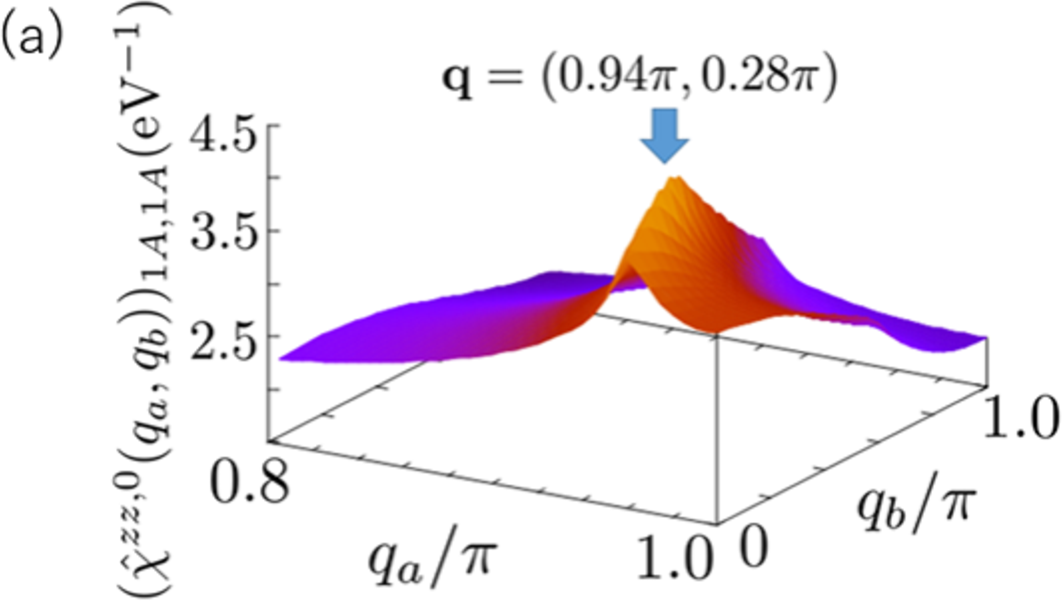}
\includegraphics[width=70mm]{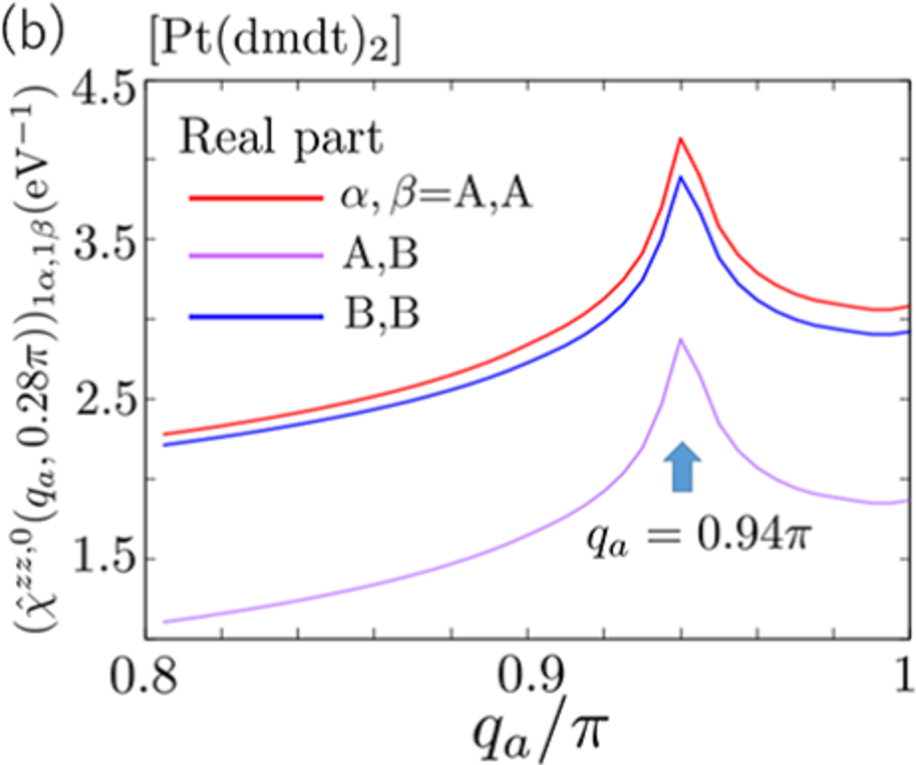}
\end{center}
\caption{(Color online) (a) Bare longitudinal susceptibility $(\hat{\chi}^{zz,0}(\textbf{q}))_{1A,1A}$ of [Pt(dmdt)$_2$] in $q_a$-$q_b$ plane. Wavenumber that results in peak corresponds to nesting vector \textbf{Q}. (b) Red, blue, and purple lines denote real parts of $(\hat{\chi}^{zz,0}(\textbf{q}))_{1A,1A}$, $(\hat{\chi}^{zz,0}(\textbf{q}))_{1A,1B}$, and $(\hat{\chi}^{zz,0}(\textbf{q}))_{1B,1B}$, respectively. 
}
\label{chi0_nd}
\end{figure}

We calculate the wavenumber dependence of the bare longitudinal susceptibility using Eq.~(14). In the absence of SOC, the transverse spin susceptibility is equivalent to the longitudinal spin susceptibility because of SU(2) symmetry. Figure~\ref{chi0_nd}(a) illustrates the bare longitudinal susceptibility $(\hat{\chi}^{zz,0}(\textbf{q}))_{1A,1A}$ of [Pt(dmdt)$_2$] in the $q_a$-$q_b$ plane. It has a peak at $(q_a,q_b)$=$(0.94\pi,0.28\pi)$, which corresponds to the nesting vector \textbf{Q} in Fig.~\ref{FS_nd}(a). Moreover, Fig.~\ref{chi0_nd}(b) shows the real parts of $(\hat{\chi}^{zz,0}(\textbf{q}))_{1A,1A}$, $(\hat{\chi}^{zz,0}(\textbf{q}))_{1A,1B}$, and $(\hat{\chi}^{zz,0}(\textbf{q}))_{1B,1B}$, where the horizontal axis is $q_a$. The wavenumber $q_b$ is fixed at $0.28\pi$. Because the real parts of $(\hat{\chi}^{zz,0}(\textbf{q}))_{1A,1B}$ are positive, the directions of the spins of orbits A and B at edge $i$=$1$ are the same. For [Ni(dmdt)$_2$], a similar result is obtained. The bare longitudinal susceptibility of [Ni(dmdt)$_2$] at the edge has a peak at $(q_a,q_b)$=$(0.95\pi,0.22\pi)$. 

\begin{figure}[htpb]
\begin{center}
\includegraphics[width=70mm]{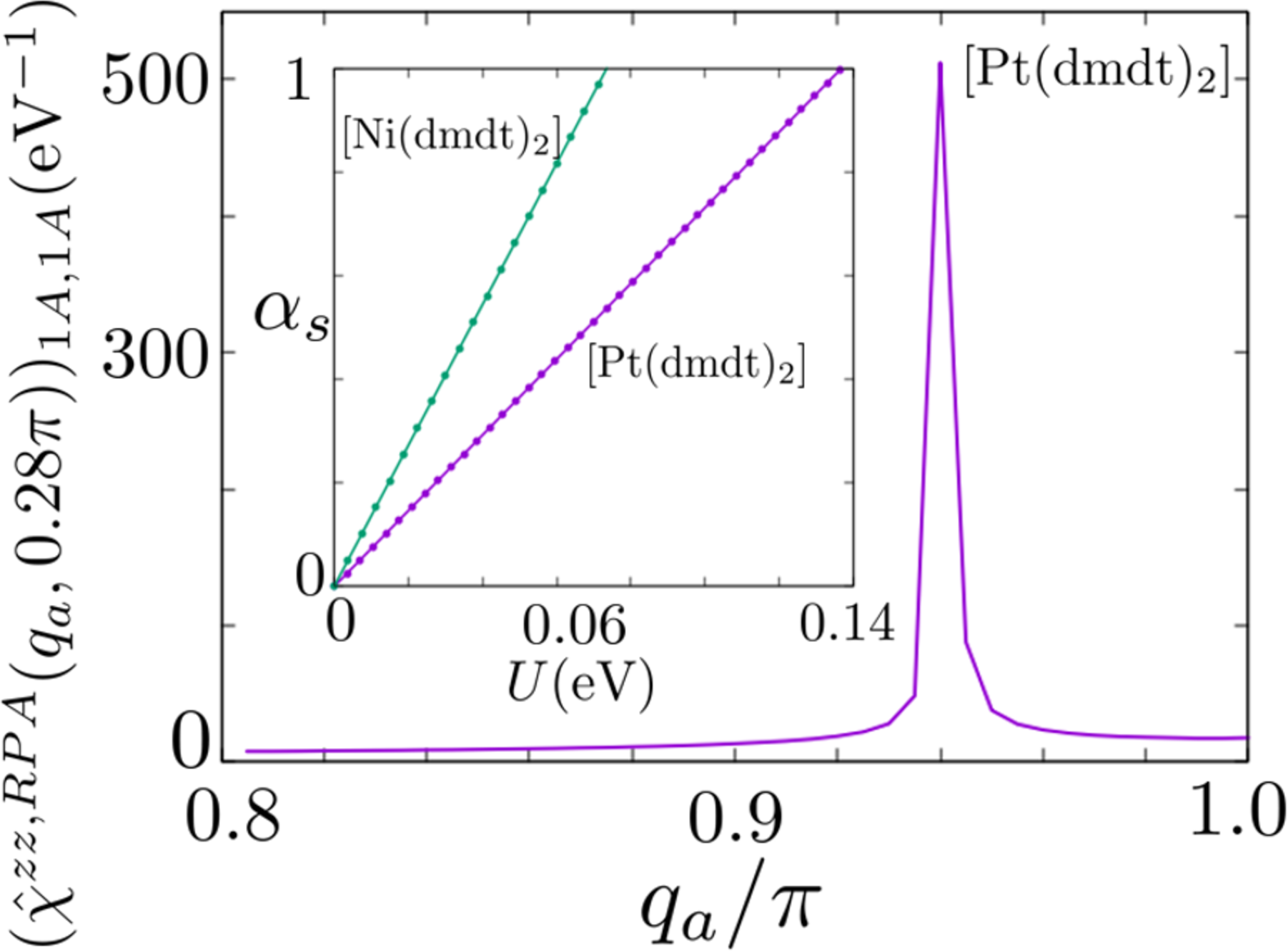}
\end{center}
\caption{(Color online) 
Longitudinal spin susceptibility $(\hat{\chi}^{zz,RPA}(q_a,0.28\pi))_{1A,1A}$ of [Pt(dmdt)$_2$], where the horizontal axis is $q_a$. Furthermore, $U$=$0.135$ and $\alpha_s$=$0.98$. Longitudinal spin fluctuation is enhanced at nesting vector $\textbf{Q}$. The inset shows the $U$ dependence of the Stoner factor. The Stoner factors of [Pt(dmdt)$_2$] and [Ni(dmdt)$_2$] are denoted by purple and green lines, respectively. $U$=$0.137$ results in [Pt(dmdt)$_2$] $\alpha_s$=$1$, and $U$=$0.073$ results in [Ni(dmdt)$_2$] $\alpha_s$=$1$.}
\label{chiS_nd}
\end{figure}

In the presence of the interaction, using Eq.~(15), we calculate the longitudinal spin susceptibility $\hat{\chi}^{zz,RPA}(\textbf{q})$.
We introduce the Stoner factor $\alpha_s$. The Stoner factor is defined as the maximum eigenvalue of the matrix $\hat{U}\hat{\chi}^{zz,0}(\textbf{q})$. The Stoner factor $\alpha_s$=$1$ gives the divergence of the longitudinal spin susceptibility, Eq.~(15), and a magnetic transition occurs. Figure~\ref{chiS_nd} shows the longitudinal spin susceptibility $(\hat{\chi}^{zz,RPA}(q_a,0.28\pi))_{1A,1A}$ of [Pt(dmdt)$_2$], where the horizontal axis is $q_a$. The wavenumber $k_b$ is fixed to $0.28\pi$, and $U$=$0.135$ results in $\alpha_s$=$0.98$. The longitudinal spin fluctuation is enhanced at this wavenumber, which corresponds to the nesting vector $\textbf{Q}$. The inset illustrates the $U$ dependence of the Stoner factor $\alpha_s$, where the $\alpha_s$ values of the [Pt(dmdt)$_2$] and [Ni(dmdt)$_2$] are shown in purple and green lines, respectively. In Fig.~\ref{chiS_nd}, $U$=$0.137$ results in [Pt(dmdt)$_2$] $\alpha_s$=$1$, whereas $U$=$0.073$ results in [Ni(dmdt)$_2$] $\alpha_s$=$1$. Divergence of the spin susceptibility at the incommensurate wavenumber indicates that the SDW is induced. The result reveals that longitudinal SDW is induced at the edge. The inset of Fig.~\ref{chiS_nd} shows that a smaller Coulomb interaction induces the SDW at the edge of [Ni(dmdt)$_2$] than that needed for [Pt(dmdt)$_2$].

\begin{figure}[htpb]
\begin{center}
\includegraphics[width=70mm]{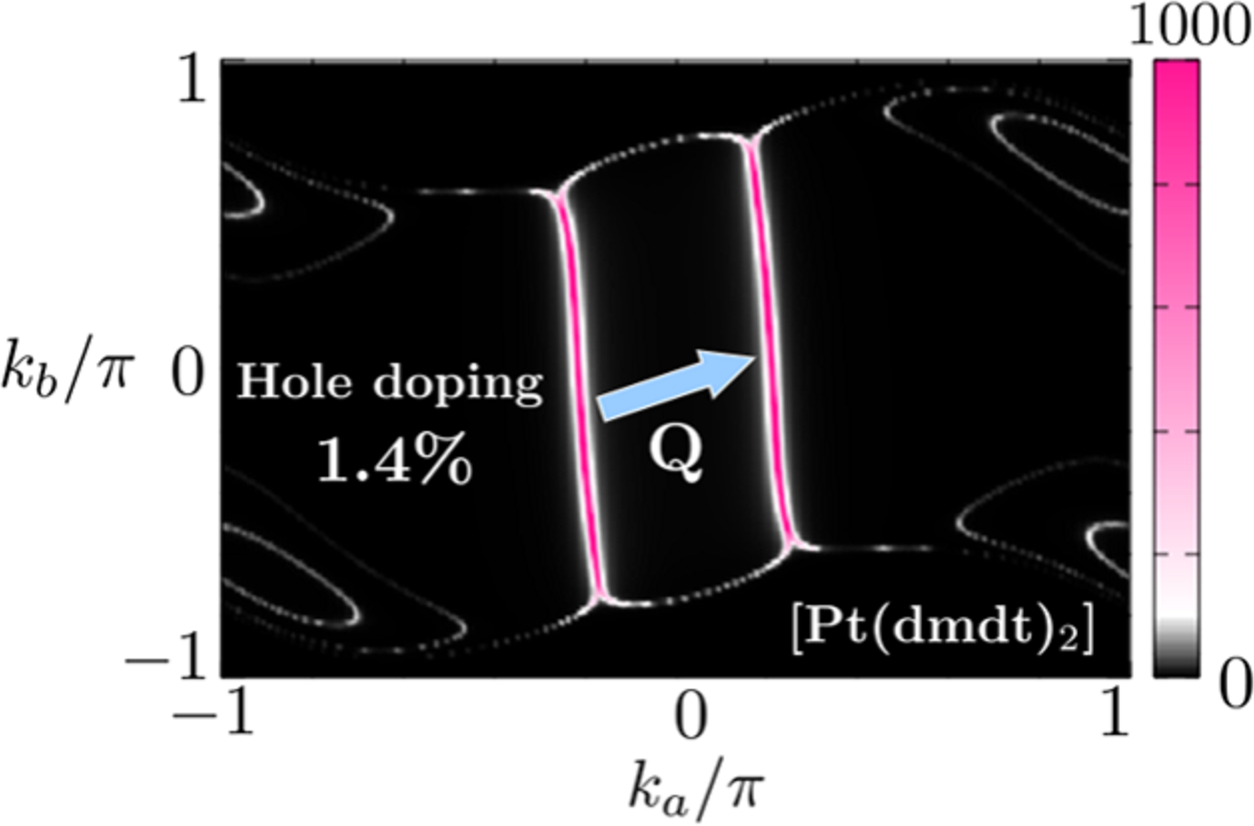}
\end{center}
\caption{(Color online) Spectral weight for the Fermi energy of [Pt(dmdt)$_2$] at the edge for $1.4\%$ hole doping, $A_{1A}(\textbf{k},0)+A_{1B}(\textbf{k},0)$. The Fermi arcs (magenta lines) become close; thus, the nesting vector \textbf{Q} becomes short. 
}
\label{FS_hd}
\end{figure}

\begin{figure}[htpb]
\begin{center}
\includegraphics[width=70mm]{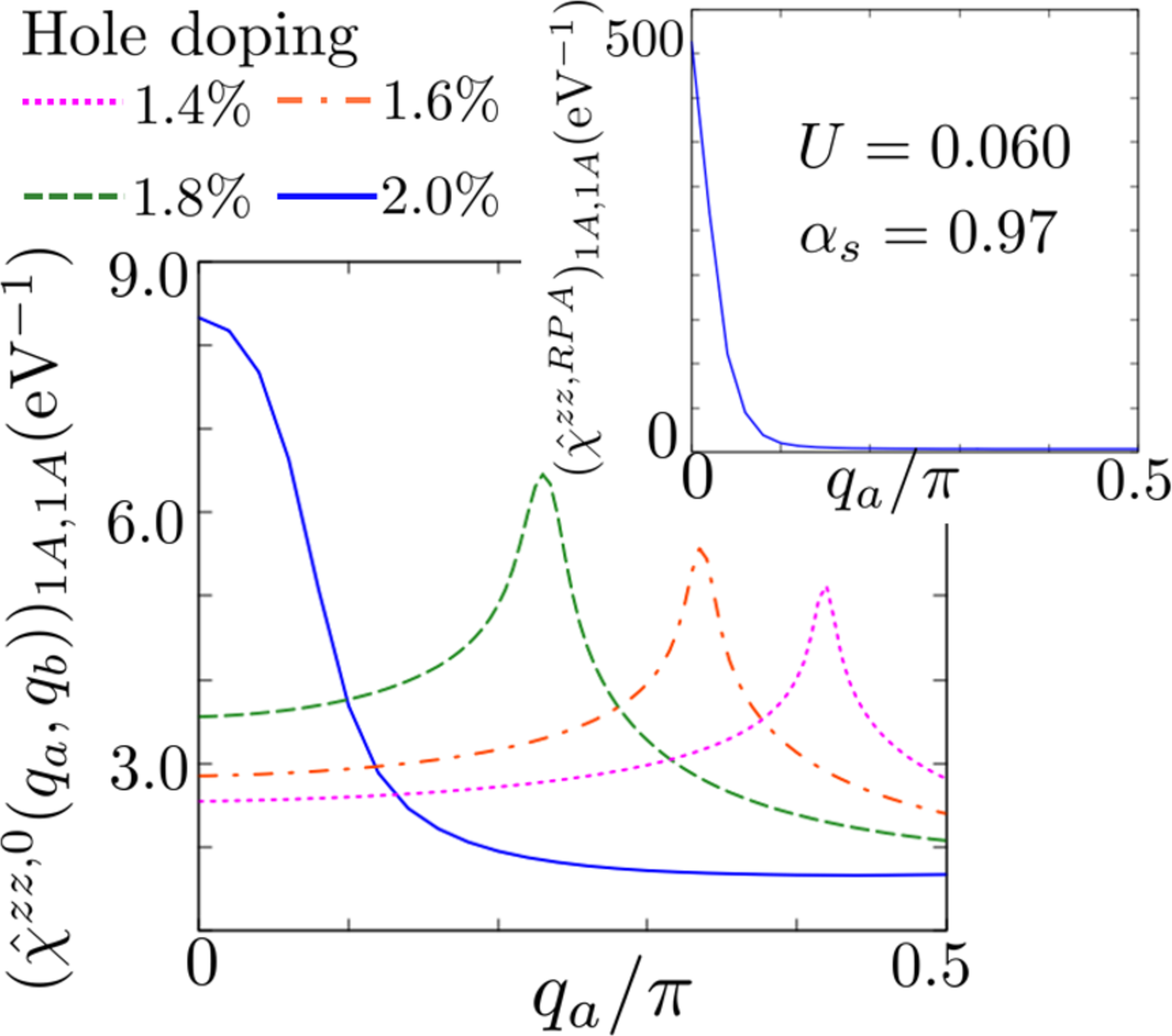}
\end{center}
\caption{(Color online) 
Bare longitudinal spin susceptibilities $(\hat{\chi}^{zz,0}(\textbf{q}))_{1A,1A}$ of [Pt(dmdt)$_2$] for hole doping. The magenta, orange, green, and blue lines represent $1.4\%$($q_b$=$0.13\pi$), $1.6\%$($q_b$=$0.13\pi$), $1.8\%$($q_b$=$0.09\pi$), and $2.0\%$($q_b$=$0$) hole doping, respectively. Hole doping changes the nesting vector $\textbf{Q}$. The inset shows longitudinal spin susceptibilities $(\hat{\chi}^{zz,RPA}(\textbf{q}))_{1A,1A}$ of [Pt(dmdt)$_2$] for $2.0\%$ hole doping, where $U$=$0.060$ and $\alpha_s=0.97$. 
}
\label{chi0_hd}
\end{figure}

We then investigate the effect of carrier doping, which modulates the Fermi arc at the edge. We calculate the Fermi arc at the edge of [Pt(dmdt)$_2$] for hole doping. Figure~\ref{FS_hd} shows the spectral weight for the Fermi energy of [Pt(dmdt)$_2$] at the edge for $1.4\%$ hole doping. Two Fermi arcs move to $k_a$=$0$ for hole doping, whereas the Fermi arc moves to $k_a$=$\pm\pi$ for electron doping. Because of the modulation of the Fermi arc, the nesting vector $\textbf{Q}$ changes. Figure~\ref{chi0_hd} visualizes the bare longitudinal spin susceptibilities $(\hat{\chi}^{zz,0}(\textbf{q}))_{1A,1A}$ of [Pt(dmdt)$_2$] for hole doping, where the horizontal axis represents $q_a$. The rates of hole doping are $1.4$, $1.6$, $1.8$, and $2.0\%$. As the rate of doping increases, the $q_a$ component of the nesting vector approaches $0$. The fixed wavenumber $q_b$ is determined such that the bare spin susceptibility attains the maximum values. The inset shows the longitudinal spin susceptibility $(\hat{\chi}^{zz,RPA}(q_a,0))_{1A,1A}$ of [Pt(dmdt)$_2$] for $2.0\%$ hole doping. $U$=$0.060$ results in a Stoner factor of $\alpha_s$=$0.97$. 

\begin{figure}[htpb]
\begin{center}
\includegraphics[width=70mm]{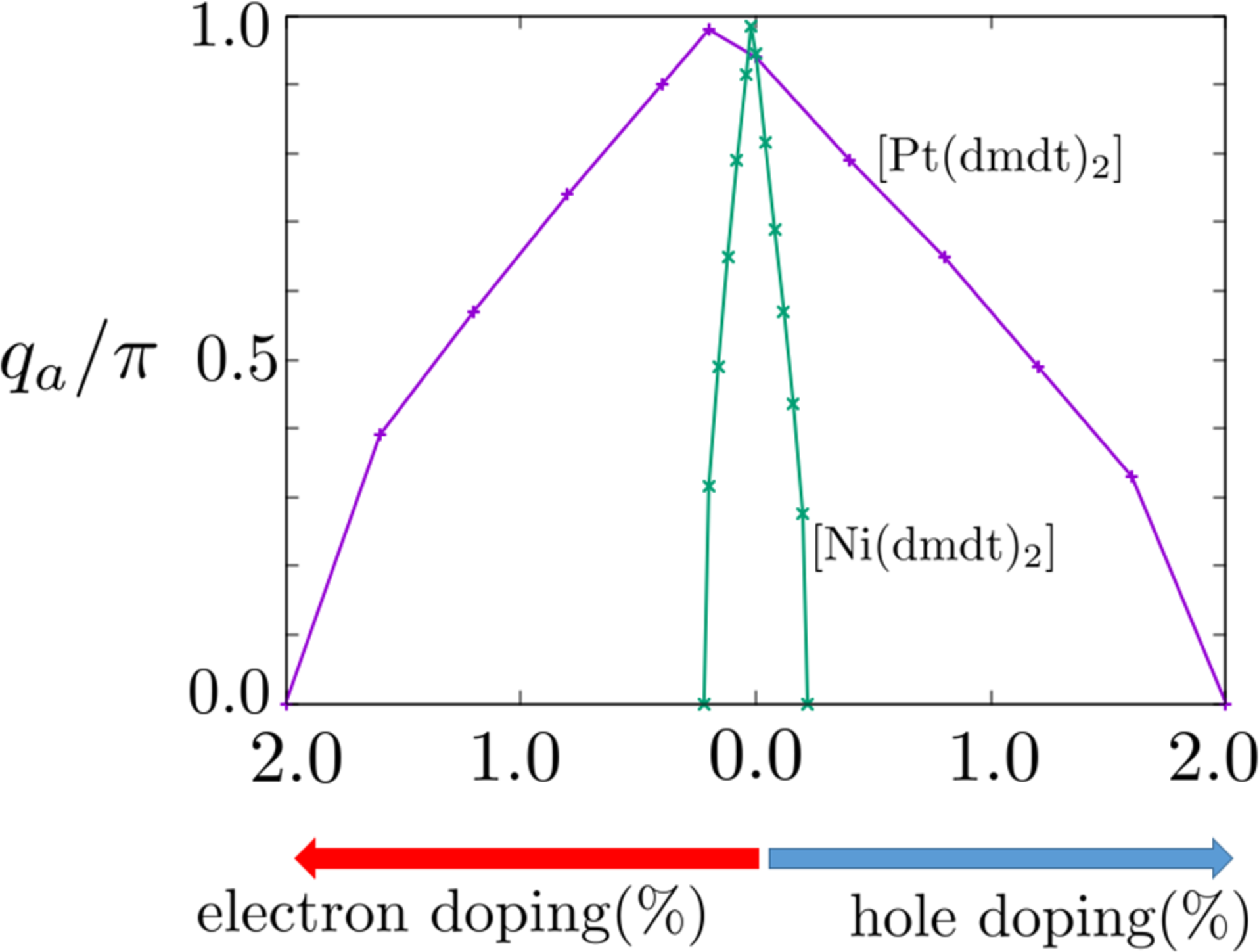}
\end{center}
\caption{(Color online) Doping dependence of the $q_a$ component of the nesting vector. The purple and green lines represent [Pt(dmdt)$_2$] and [Ni(dmdt)$_2$], respectively. 
}
\label{doping}
\end{figure}

Figure~\ref{doping} visualizes the relationship between the rate of carrier doping and the $q_a$ component of the nesting vector \textbf{Q} for both materials. For [Pt(dmdt)$_2$], $2.0\%$ hole doping and $2.0\%$ electron doping yield $\textbf{Q}$=$\textbf{0}$. For [Ni(dmdt)$_2$], $0.23\%$ hole doping and $0.23\%$ electron doping give a result equivalent to that for [Pt(dmdt)$_2$]. The nesting vector $\textbf{Q}$=$\textbf{0}$ induces edge ferromagnetism. Carrier doping can change the nesting vector $\textbf{Q}$ and control the magnetic structure at the (001) edge. 
\subsection{Presence of spin--orbit coupling}
In this section, we investigate edge magnetism in the presence of SOC. According to the first-principles calculation, the SOC constant of [Pt(dmdt)$_2$] is $\lambda$=$0.05$ ($\sim$ $0.0022$ eV for the hopping energies $t_{4}$ and $t_{5}$). In this study, we use $\lambda$=$0.05$ for [Ni(dmdt)$_2$]. For [Ni(dmdt)$_2$], $\lambda$=$0.05$ results in hopping energies $t_{4}$ and $t_{5}$ $\sim$$0.0016$ eV. Because of the SOC, a helical edge state occurs, and the Fermi arc splits. 
\begin{figure}[htpb]
\begin{center}
\includegraphics[width=70mm]{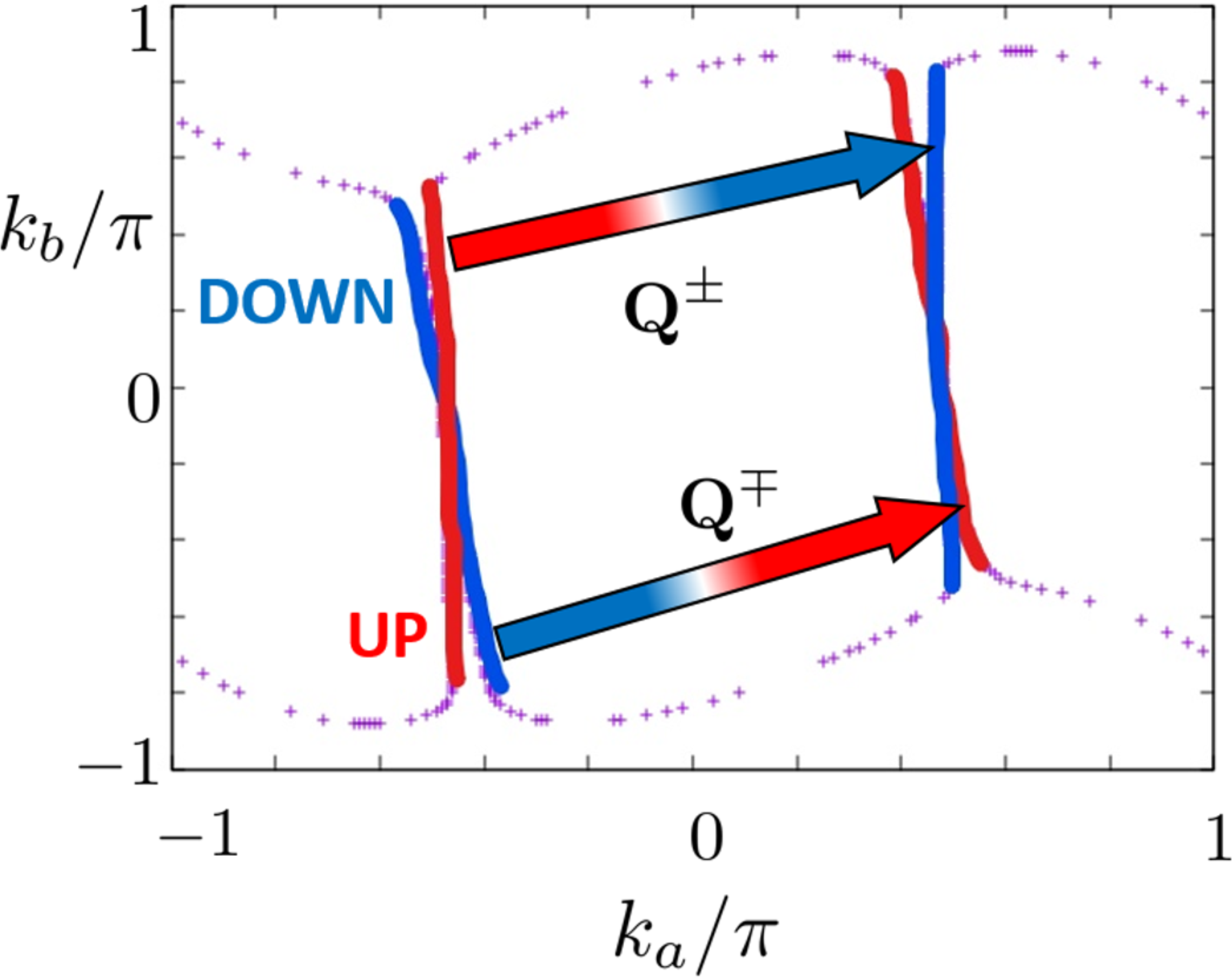}
\end{center}
\caption{(Color online) The red and blue lines represent the Fermi arcs with up and down spins, respectively, on the $i$=$1$ edge. The Fermi arc has independent nesting vectors $\textbf{Q}^{\pm}$ and $\textbf{Q}^{\mp}$.
}
\label{FS_SOC_nd}
\end{figure}

Figure~\ref{FS_SOC_nd} shows the Fermi arcs at the edge in the presence of SOC. The red and blue lines represent the Fermi arcs with up and down spins, respectively. The nesting vectors, which connect the same spin bands, degrade; however, different good nesting vectors $\textbf{Q}^{\pm}$ and $\textbf{Q}^{\mp}$ appear. $\textbf{Q}^{\pm}$ is the nesting vector that goes to the band with the down spin from the band with the up spin, whereas $\textbf{Q}^{\mp}$ is the nesting vector that goes to the band with the up spin from the band with the down spin. $\textbf{Q}^{\pm}$ and $\textbf{Q}^{\mp}$ connect the Fermi arcs of the different bands. Thus, they are excitonic. 

\begin{figure}[htpb]
\begin{center}
\includegraphics[width=70mm]{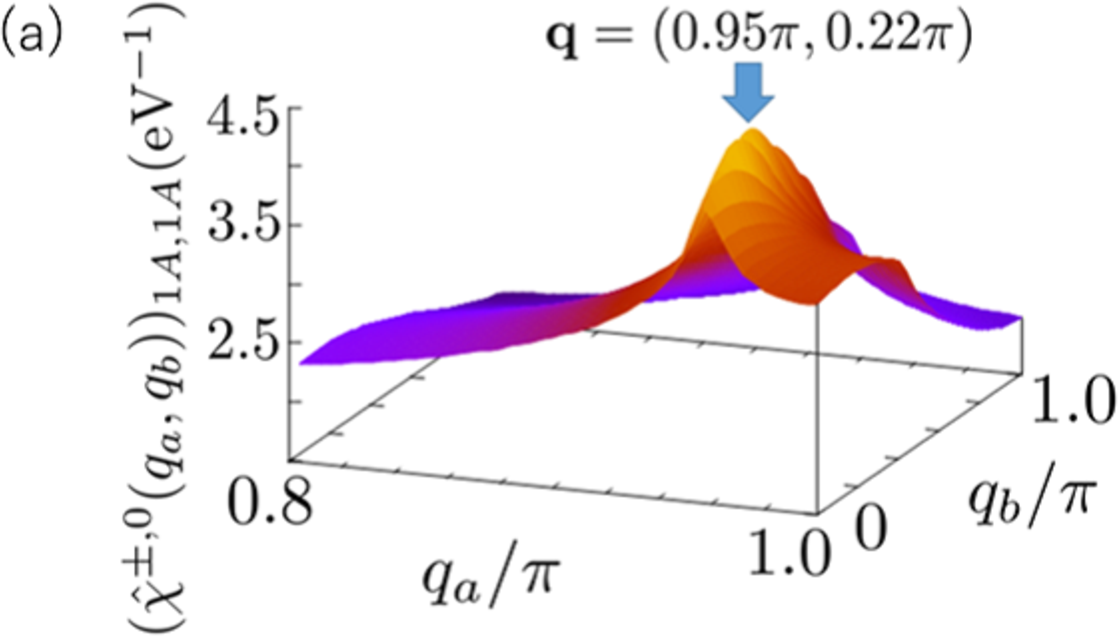}
\includegraphics[width=70mm]{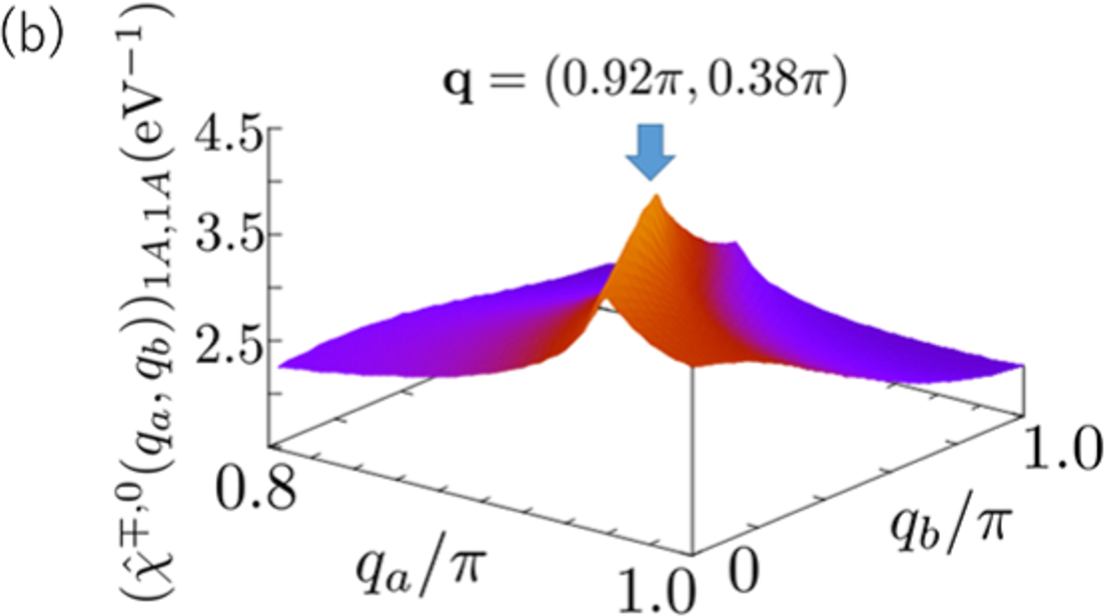}
\includegraphics[width=70mm]{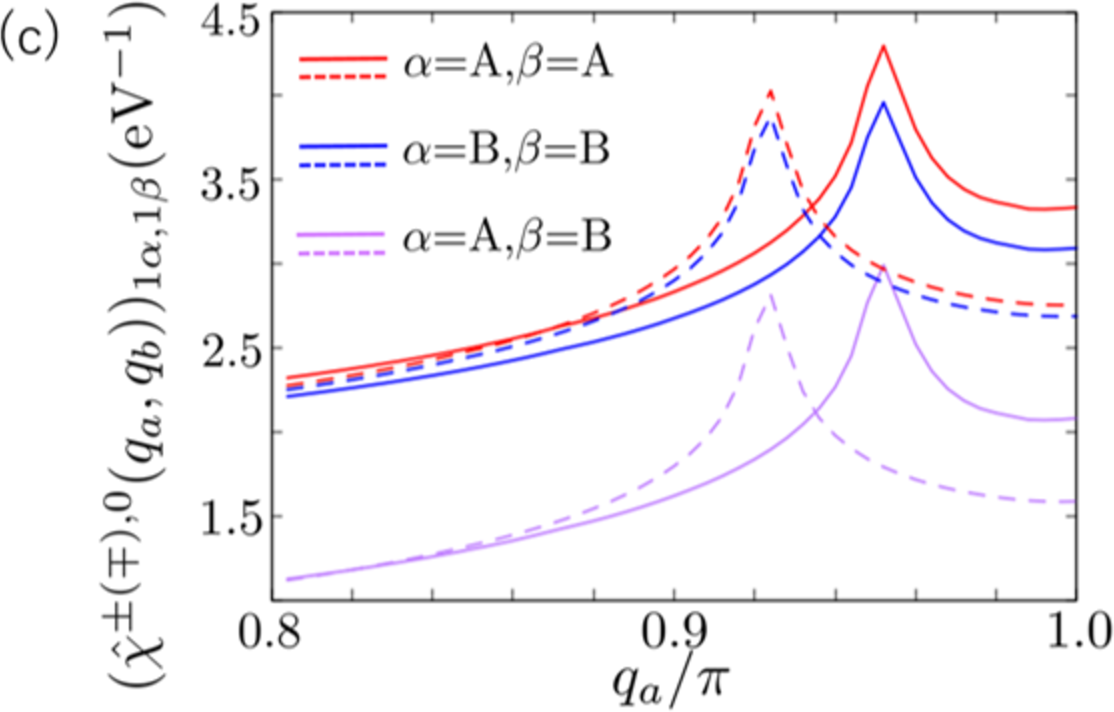}
\end{center}
\caption{(Color online) (a) $(\hat{\chi}^{\pm,0}(\textbf{q}))_{1A,1A}$ in $q_a$-$q_b$ plane. (b) $(\hat{\chi}^{\mp,0}(\textbf{q}))_{1A,1A}$. (c) $q_a$ dependence of bare transverse spin susceptibility. Solid and broken lines denote $\hat{\chi}^{\pm,0}(q_a,0.22\pi)$ and $\hat{\chi}^{\mp,0}(q_a,0.38\pi)$, respectively. Red, blue, and purple represent 
1A1A, 1B1B, and 1A1B components, respectively. Wavenumbers that result in peaks correspond to the nesting vectors $\textbf{Q}^{\pm}$ and $\textbf{Q}^{\mp}$.
}
\label{chi0_pmmp}
\end{figure}

We calculate the bare transverse spin susceptibility $\hat{\chi}^{\pm,0}(\textbf{q})$ and $\hat{\chi}^{\mp,0}(\textbf{q})$ using Eq.~(19), where it is expected that the good nesting vectors $\textbf{Q}^{\pm}$ and $\textbf{Q}^{\mp}$ enhance them.
Figure~\ref{chi0_pmmp}(a) and \ref{chi0_pmmp}(b) show the bare transverse spin susceptibility $(\hat{\chi}^{\pm,0}(\textbf{q}))_{1A,1A}$ and $(\hat{\chi}^{\mp,0}(\textbf{q}))_{1A,1A}$ of [Pt(dmdt)$_2$] in the $q_a$-$q_b$ plane. $(\hat{\chi}^{\pm,0}(\textbf{q}))_{1A,1A}$ and $(\hat{\chi}^{\mp,0}(\textbf{q}))_{1A,1A}$ have peaks at incommensurate wavenumbers $(q_a,q_b)$=$(0.95\pi,0.22\pi)$ and $(q_a,q_b)$=$(0.92\pi,0.38\pi)$, which correspond to the nesting vectors $\textbf{Q}^{\pm}$ and $\textbf{Q}^{\mp}$, respectively. Figure~\ref{chi0_pmmp}(c) shows the real parts of $\hat{\chi}^{\pm,0}(q_a,0.22\pi)$ (solid line) and $\hat{\chi}^{\mp,0}(q_a,0.38\pi)$ (broken line), where the horizontal axis is $q_a$. The 1A1A, 1B1B, and 1A1B components are represented by the red, blue, and purple lines, respectively. Because the real parts of the 1A1B component are positive, the directions of the spins of orbits A and B at edge $i$=$1$ are the same. 
The results for [Ni(dmdt)$_2$] are similar to those for [Pt(dmdt)$_2$]. The bare transverse spin susceptibilities $(\hat{\chi}^{\pm,0}(\textbf{q}))_{1A,1A}$ and $(\hat{\chi}^{\mp,0}(\textbf{q}))_{1A,1A}$ have peaks at incommensurate wavenumbers $(q_a,q_b)$=$(0.96\pi,0.28\pi)$ and $(q_a,q_b)$=$(0.93\pi,0.25\pi)$, respectively. 

We then calculate the Stoner factors $\alpha_s^{\pm}$ and $\alpha_s^{\mp}$ at the $i$=$1$ edge. They are defined as the maximum eigenvalues of the matrices $\hat{U}\hat{\chi}^{\pm,0}(\textbf{q})$ and $\hat{U}\hat{\chi}^{\mp,0}(\textbf{q})$, where we use only the matrix elements of transverse spin susceptibility, which correspond to a few unit cells near the $i$=$1$ edge. Of the two Stoner factors, $\alpha_s^{\pm}$=$1$ causes the transverse spin susceptibility Eq.~(20) to diverge, whereas $\alpha_s^{\pm}$=$1$ causes Eq.~(21) to diverge.
\begin{figure}[htpb]
\begin{center}
\includegraphics[width=60mm]{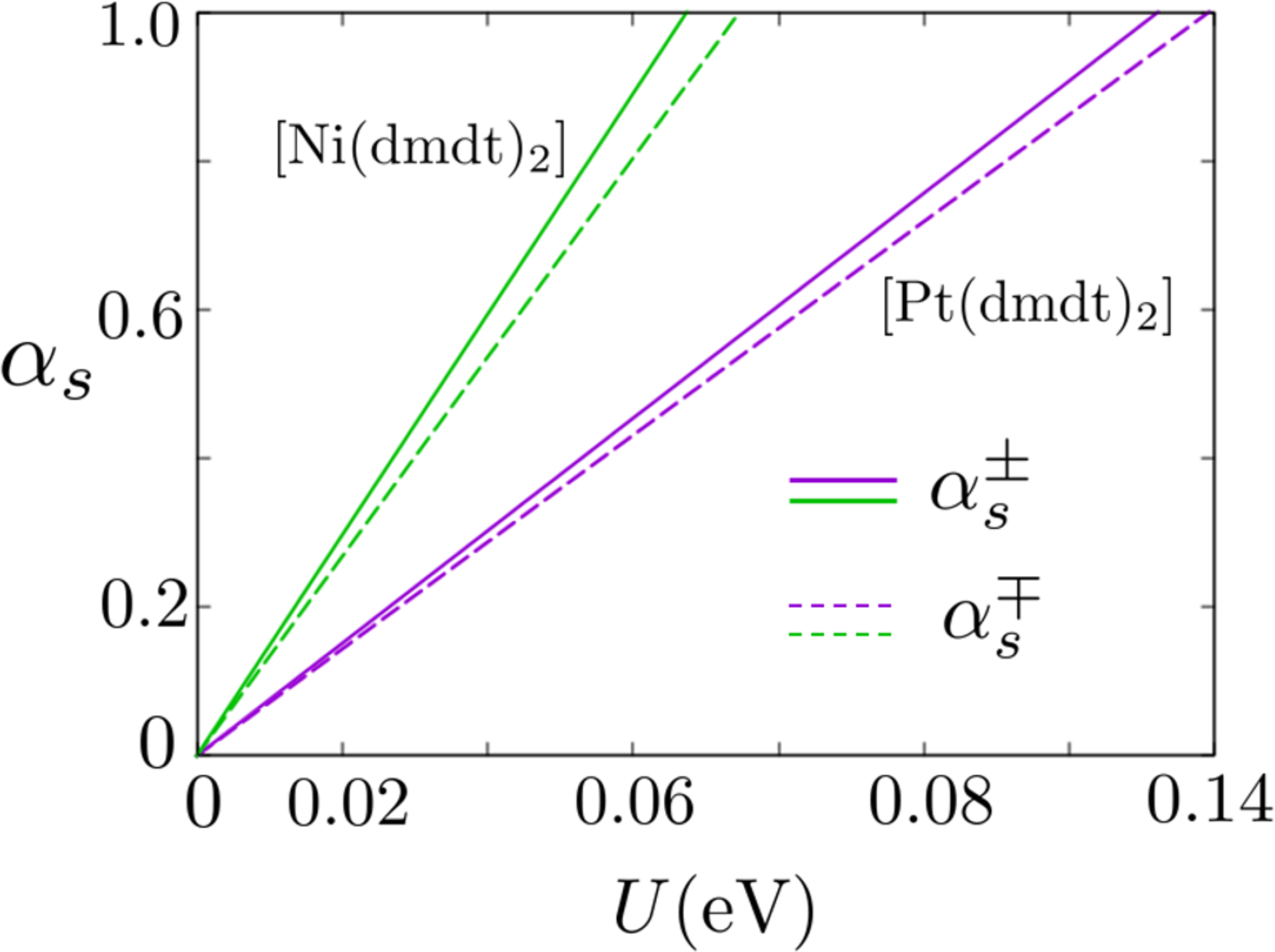}
\end{center}
\caption{(Color online) $U$ dependence of Stoner factors at $i$=$1$ edge of [Pt(dmdt)$_2$] and [Ni(dmdt)$_2$]. The purple solid and broken lines represent $\alpha_s^{\pm}$ and $\alpha_s^{\mp}$ of [Pt(dmdt)$_2$], respectively. The green solid and broken lines represent $\alpha_s^{\pm}$ and $\alpha_s^{\mp}$ of [Ni(dmdt)$_2$], respectively. 
} 
\label{chiS_pmmp}
\end{figure}
Figure~\ref{chiS_pmmp} shows the $U$ dependence of the Stoner factors $\alpha^{\pm}_{s}$ and $\alpha^{\mp}_{s}$ at the $i$=$1$ edge of [Pt(dmdt)$_2$] and [Ni(dmdt)$_2$]. The purple and green lines represent [Pt(dmdt)$_2$] and [Ni(dmdt)$_2$], whereas the solid and broken lines represent $\alpha_s^{\pm}$ and $\alpha_s^{\mp}$, respectively. At the edge of [Pt(dmdt)$_2$], $U$=$0.132$ and $U$=$0.140$ yield $\alpha_s^{\pm}$=$1$ and $\alpha_s^{\mp}$=$1$, respectively. Meanwhile, at the edge of [Ni(dmdt)$_2$], $U$=$0.067$ and $U$=$0.075$ give $\alpha_s^{\pm}$=$1$ and $\alpha_s^{\mp}$=$1$, respectively. Thus, at the $i$=$1$ edge of the two materials, an excitonic transverse SDW corresponding to $\hat{\chi}^{\pm}(\textbf{q})$ is induced. On the other hand, an excitonic transverse SDW corresponding to $\hat{\chi}^{\mp}(\textbf{q})$ is induced at the $i$=$N_c$ edge. This is because the combination of spins and energy bands at the $i$=$N_c$ edge is opposite to that at the $i$=$1$ edge. 

\begin{figure}[htpb]
\includegraphics[width=70mm]{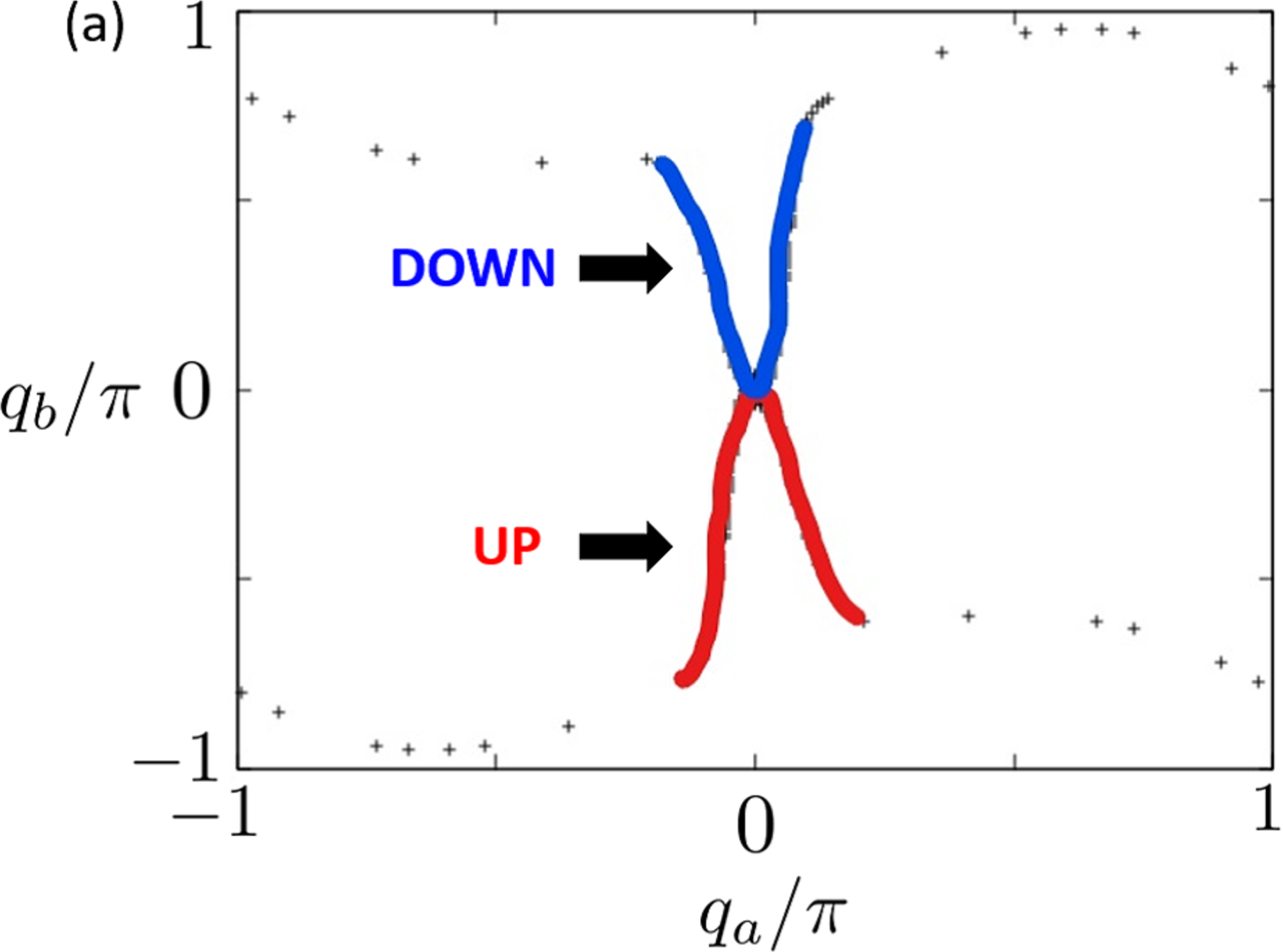}
\includegraphics[width=70mm]{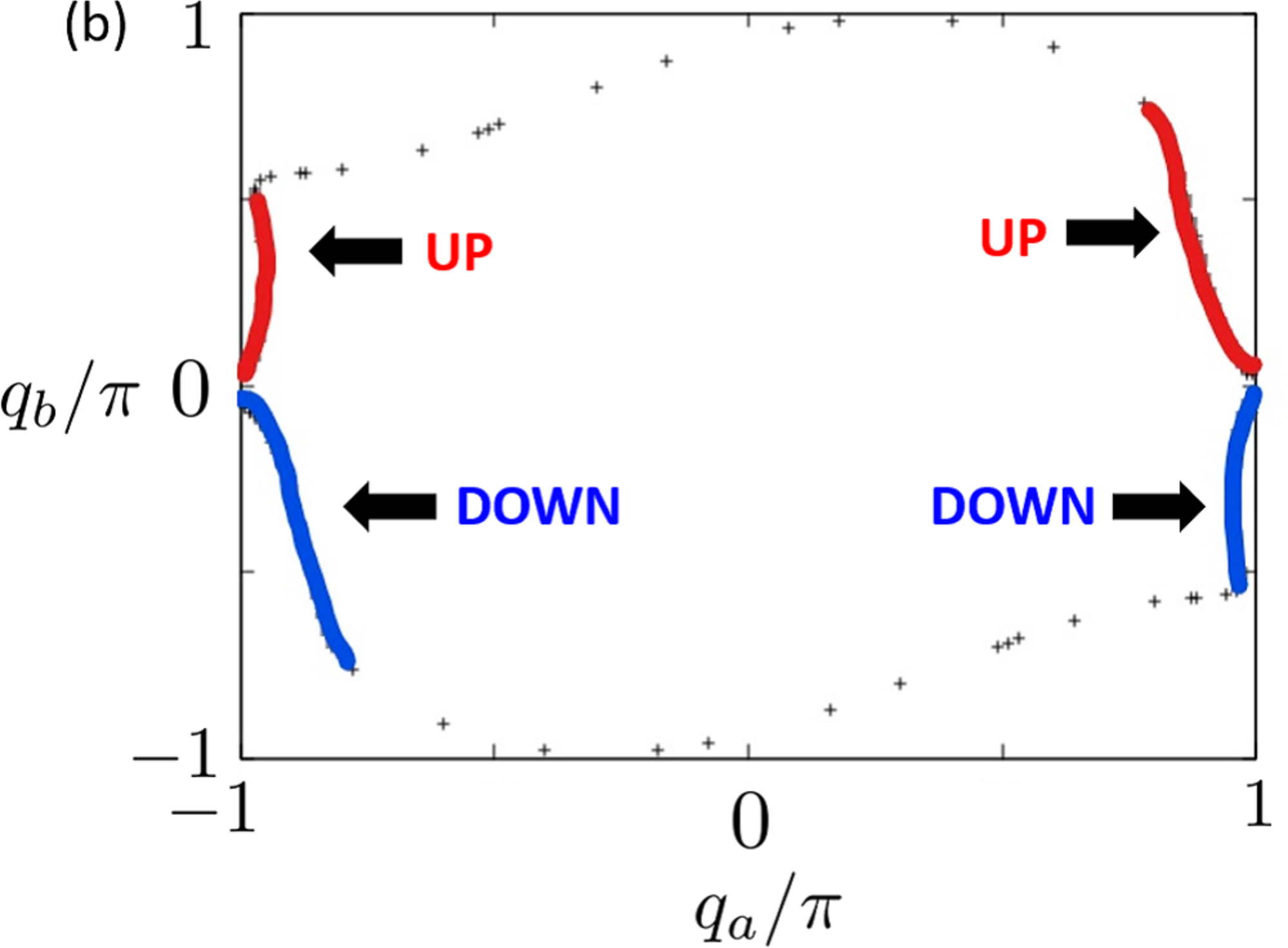}
\caption{(Color online) (a) Fermi arc for $2.0\%$ hole doping in presence of SOC. (b) Fermi arc for $2.0\%$ electron doping in presence of SOC. The red and blue lines represent Fermi arcs with up and down spins, respectively. 
}
\label{FS_SOC_dop}
\end{figure}

\begin{figure}[htpb]
\begin{center}
\includegraphics[width=70mm]{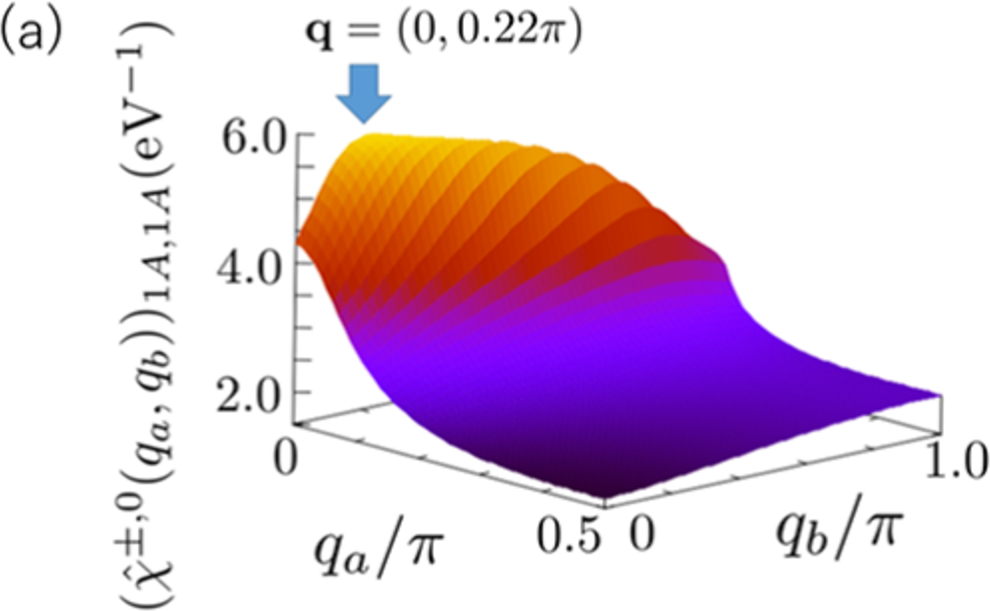}
\includegraphics[width=70mm]{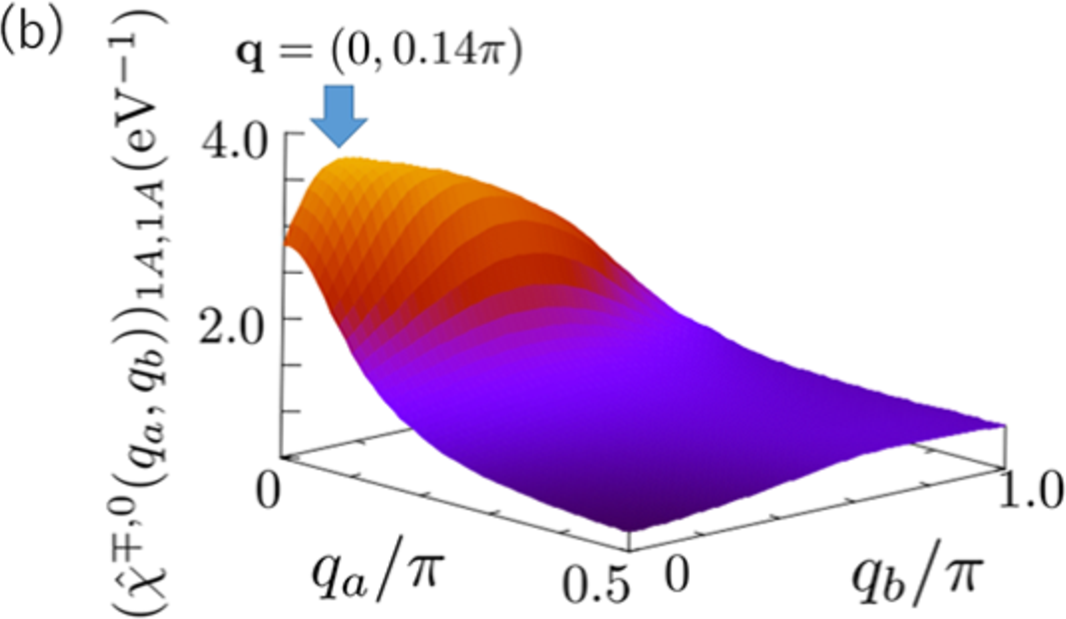}
\end{center}
\caption{(Color online) (a) Bare transverse spin susceptibility $(\hat{\chi}^{\pm,0}(q_a,q_b))_{1A,1A}$ for $2.0\%$ hole doping in $q_a$-$q_b$ plane. (b) $(\hat{\chi}^{\mp,0}(q_a,q_b))_{1A,1A}$ for $2.0\%$ electron doping in $q_a$-$q_b$ plane. 
}
\label{chi0_dop_SOC}
\end{figure}

$\textbf{Q}^{\pm}$ at the $i$=$1$ edge is equal to $\textbf{Q}^{\mp}$ at the $i$=$N_c$ edge because of time reversal symmetry. In the presence of SOC, carrier doping modulates the Fermi arc as also observed in the absence of SOC. However, the Fermi arc in the presence of SOC is different from that in the absence of SOC. We calculate the Fermi arc for carrier doping in the presence of SOC. Figures~\ref{FS_SOC_dop}(a) and \ref{FS_SOC_dop}(b) show the Fermi arcs of [Pt(dmdt)$_2$] in the presence of SOC for $2.0\%$ hole doping and $2.0\%$ electron doping, respectively. The red and blue lines represent Fermi arcs with up and down spins, respectively. The black points denote the energy dispersion of the bulk, which does not play an important role in this study. Calculation of the bare transverse spin susceptibility reflects the modulation of the Fermi arc. Figure~\ref{chi0_dop_SOC}(a) shows the transverse spin susceptibility $(\hat{\chi}^{\pm,0}(q_a,q_b))_{1A,1A}$ of [Pt(dmdt)$_2$] for $2.0\%$ hole doping in the $q_a$-$q_b$ plane, whereas Fig.~\ref{chi0_dop_SOC}(b) illustrates $(\hat{\chi}^{\mp,0}(q_a,q_b))_{1A,1A}$ for $2.0\%$ electron doping in the $q_a$-$q_b$ plane, which have maximum values at $(q_a,q_b)$=$(0,0.22\pi)$ and $(q_a,q_b)$=$(0,0.14\pi)$, respectively. For [Ni(dmdt)$_2$], we obtain results similar to those for [Pt(dmdt)$_2$]. The transverse spin susceptibilities $(\hat{\chi}^{\pm,0}(q_a,q_b))_{1A,1A}$ for $0.23\%$ hole doping and $(\hat{\chi}^{\mp,0}(q_a,q_b))_{1A,1A}$ for $0.23\%$ electron doping have maximum values at $(q_a,q_b)$=$(0.11\pi,0.24\pi)$ and $(q_a,q_b)$=$(0,0.16\pi)$, respectively. As in the absence of SOC, carrier doping modulates the Fermi arc and magnetic structure at the (001) edge in the presence of SOC. 

\begin{figure}[htpb]
\begin{center}
\includegraphics[width=40mm]{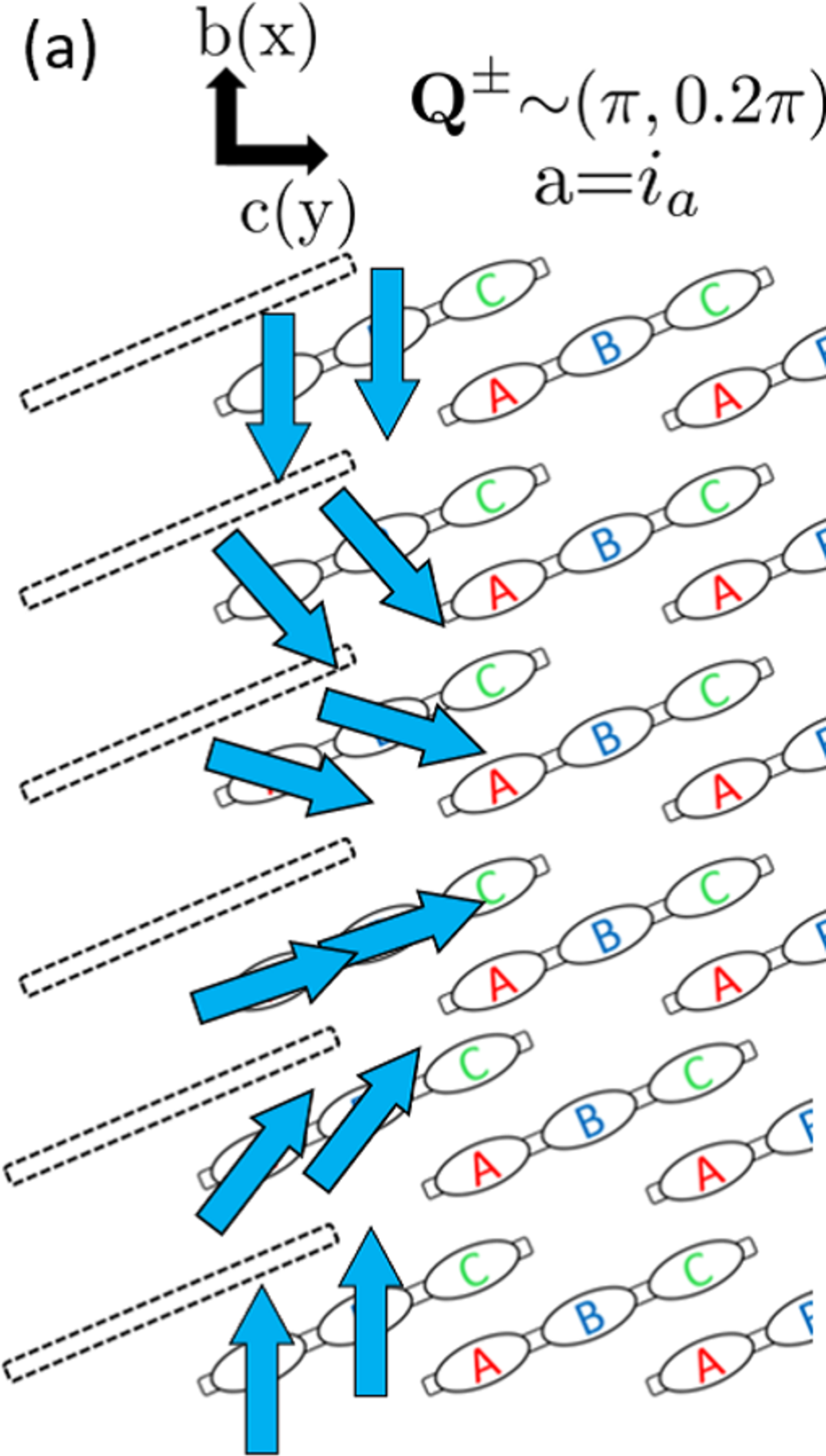}\\
\includegraphics[width=40mm]{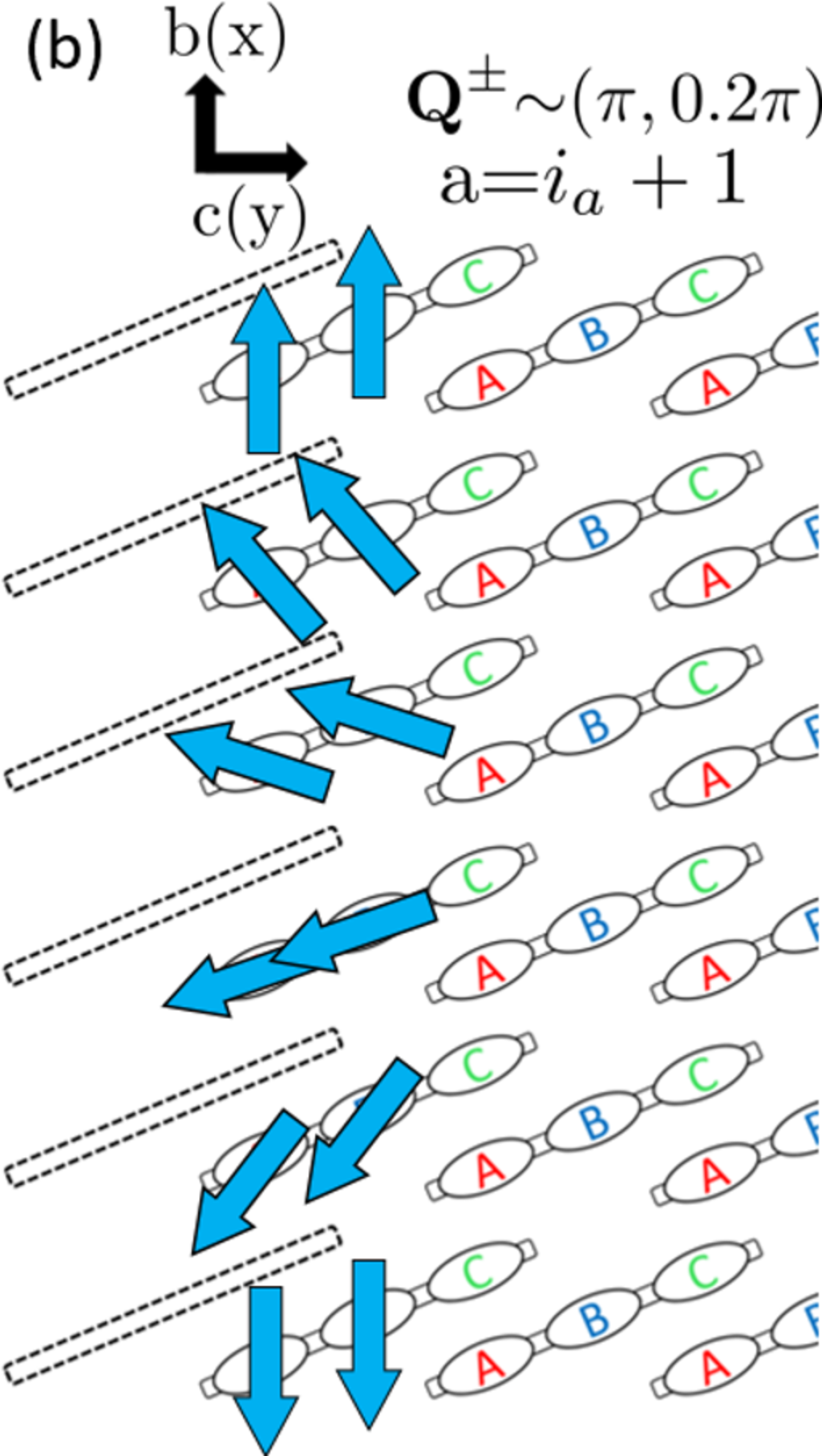}
\end{center}
\caption{(Color online) Spin structure at (001) edge of [Pt(dmdt)$_2$] in the presence of SOC. (a) and (b) are on the $a$=$i_a$ and $a$=$i_a+1$ layers, respectively.
}
\label{spin_structure}
\end{figure}

Finally, we schematically show the spin structure at the (001) edge of [Pt(dmdt)$_2$] in the presence of the SOC. At the $i$=$1$ edge, the SDW corresponding to the transverse spin susceptibility $\hat{\chi}^{\pm}(\textbf{Q}^{\pm})$ is induced. The transverse SDW represents the rotation of the spins in the $b$-$c$ plane because we select the $a$-axis as the quantization axis in the present paper. Figures~\ref{spin_structure}(a) and \ref{spin_structure}(b) show the transverse SDW at the $i$=$1$ edge on the $a$=$i_a$ and $a$=$i_a+1$ layers, respectively, in the absence of carrier doping. In Figs.~\ref{spin_structure}(a) and \ref{spin_structure}(b), we approximate the wavenumber of SDW $\textbf{Q}^{\pm}$$\sim$$(\pi,0.2\pi)$, which represents antiferromagnetic spin rotation along the $a$-axis and ten times periodic spin rotation along the $b$-axis. Meanwhile, at the $i$=$N_c$ edge, the spins reversely rotate and the period of the rotation is the same as the $i$=$1$ edge. This indicate the helical SDW\cite{Jiang2011}, which results from the SOC and breaking of spatial inversion symmetry at the edge. 

\section{Summary and Discussion}
In this study, we constructed three-orbital tight-binding models in the presence of SOC, which describe the electrical states of [Pt(dmdt)$_2$] and [Ni(dmdt)$_2$] in a unified manner using the first-principles calculation packages Quantum ESPRESSO and RESPACK. 
We determined that [Ni(dmdt)$_2$] is a Dirac nodal line system that is similar to [Pt(dmdt)$_2$] but closer to a two-dimensional system.
By applying tight-binding models to cylindrical systems with (001) edges, we analyzed the properties of the edge states. In the absence of SOC, the edge state, which is characterized by quasi-one-dimensional energy dispersion, emerges between the Dirac nodal lines to the (001) direction.
We observed an incommensurate nesting vector $\textbf{Q}$ between the Fermi arcs of the edge state. Meanwhile, in the presence of SOC, helical edge states emerge with two different incommensurate nesting vectors $\textbf{Q}^{\pm}$ and $\textbf{Q}^{\mp}$ because of the splitting of Fermi arcs between up and down spins. This helical split is due to SOC and the breaking of spatial inversion symmetry at the edge. 

To investigate the edge magnetism induced by such nesting vectors, we calculated the longitudinal and transverse spin susceptibilities, by real-space-dependent RPA for the three-orbital Hubbard model describing [Pt(dmdt)$_2$] and [Ni(dmdt)$_2$]. In the absence of SOC, the nesting vector $\textbf{Q}$ enhances the longitudinal spin susceptibility at the (001) edge. 
Thus, longitudinal SDW occurs at the (001) edge. We also investigated the effect of carrier doping. Carrier doping modulates the Fermi arc, and the nesting vector $\textbf{Q}$ varies. In particular, we observed that hole doping and electron doping at $2.0\%$ for [Pt(dmdt)$_2$] and $0.23\%$ for [Ni(dmdt)$_2$] induce edge ferromagnetism. In the presence of SOC, the nesting vectors $\textbf{Q}^{\pm}$ and $\textbf{Q}^{\mp}$ induce transverse SDW at the edge. In addition to what happens in the absence of SOC, the carrier doping modulates the Fermi arcs, and the nesting vectors $\textbf{Q}^{\pm}$ and $\textbf{Q}^{\mp}$ vary. Therefore, the magnetic structure at the (001) edge is extremely sensitive to carrier doping. Figure \ref{chi0_uc} shows that the bare longitudinal spin susceptibility markedly decays from the edge to the bulk. 
The effect of the edge state reaches within a few unit cells near the edge. 
Hence, it is considered that the large magnetic moment is induced only near the edge. 
Thus, we consider that the topological property in the bulk is not affected by the edge SDW.

[Pt(dmdt)$_2$] and [Ni(dmdt)$_2$] would have become weak topological insulators if the spin--orbit interaction was sufficiently large.\cite{T.Kawamura2020} Thus, these can be positioned as a new variation of magnetic order at the edges of topological insulators.\cite{Jiang2011}
In the experiment, it is predicted that the edge SDW is observed as a magneto-optical Kerr effect. Because SOC is finite in our reality, transverse SDW can be experimentally observed at the (001) edge. Such optical experiments are expected to provide information on small amounts of carrier doping and spin--orbit interaction. For organic conductors, carriers can be strictly controlled in the experiment; small amounts of carrier doping have been demonstrated to be controllable in organic conductors.
For example, a field effect transistor (FET) can change amounts of carriers and modulate the Fermi arc,\cite{TajimaFET,YamamotoFET} whereas
transport properties such as the hole coefficient are sensitive probes for carrier doping.\cite{N.Tajima2012,A.Kobayashi2008}
The sensitivity of SDW for carrier doping can also be useful as a probe for controlling extremely dilute carriers. \\



\begin{acknowledgment}

The authors would like to thank K. Yada for fruitful discussion. 
This work was supported by MEXT (JP) JSPJ (Grants No. 15K05166, 19H01846, and 17K05846). The computation in this work was performed using the facilities of the Supercomputer Center, Institute for Solid State Physics, University of Tokyo.

\end{acknowledgment}

\appendix
\section{Six Orbital Tight-Binding Model of [Ni(dmdt)$_2$]}

\begin{figure}[htpb]
\begin{center}
\includegraphics[width=70mm]{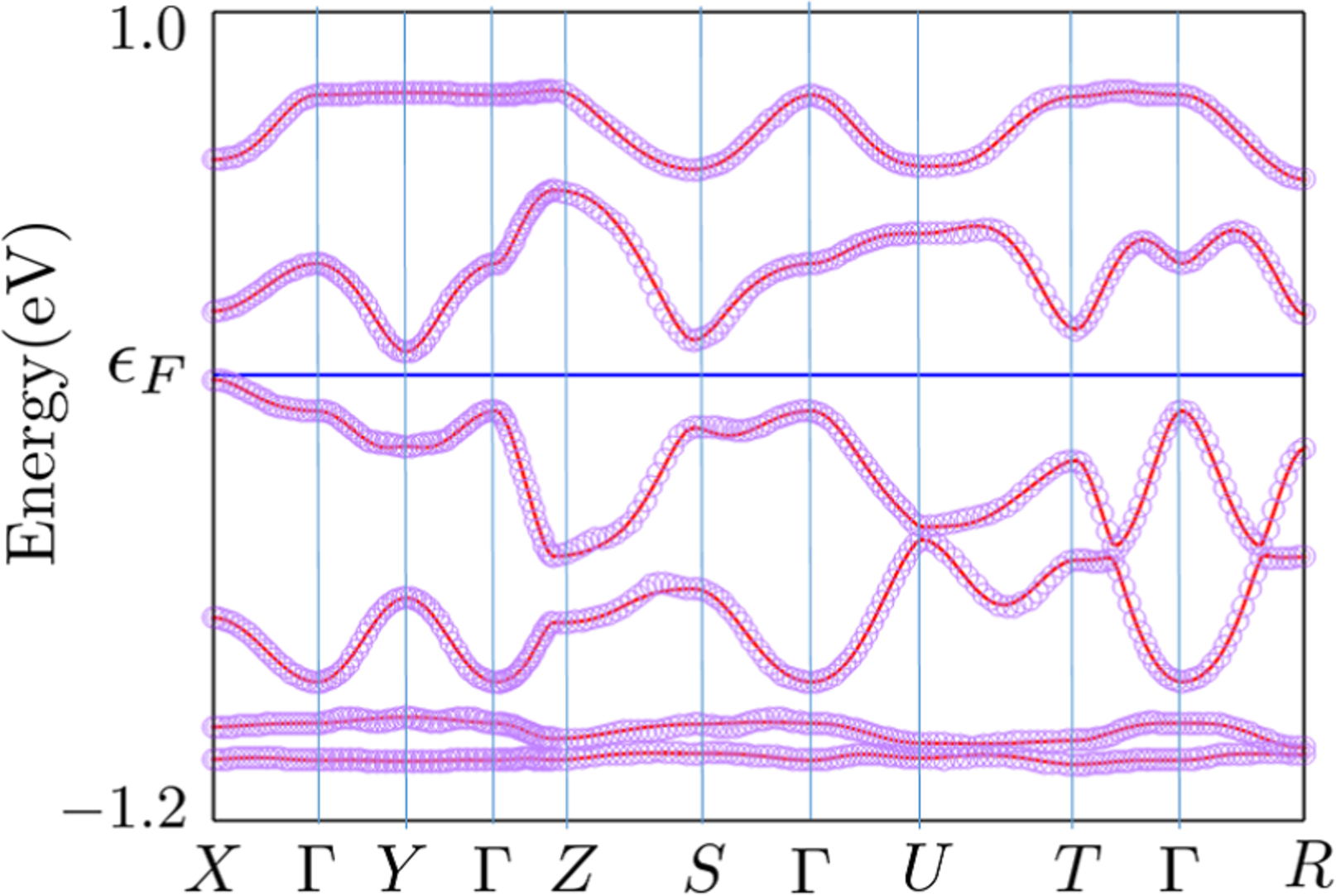}
\end{center}
\caption{(Color online) Red lines and purple circles depict energy band structures of [Ni(dmdt)$_2$] obtained via first-principles calculation and Wannier fitting, respectively. Horizontal axis represents connecting symmetry points. Wannier fitting reproduces energy band structure calculated via first-principles calculation. 
} 
\label{wannier_6}
\end{figure}

\begin{figure}[htpb]
\begin{center}
\includegraphics[width=60mm]{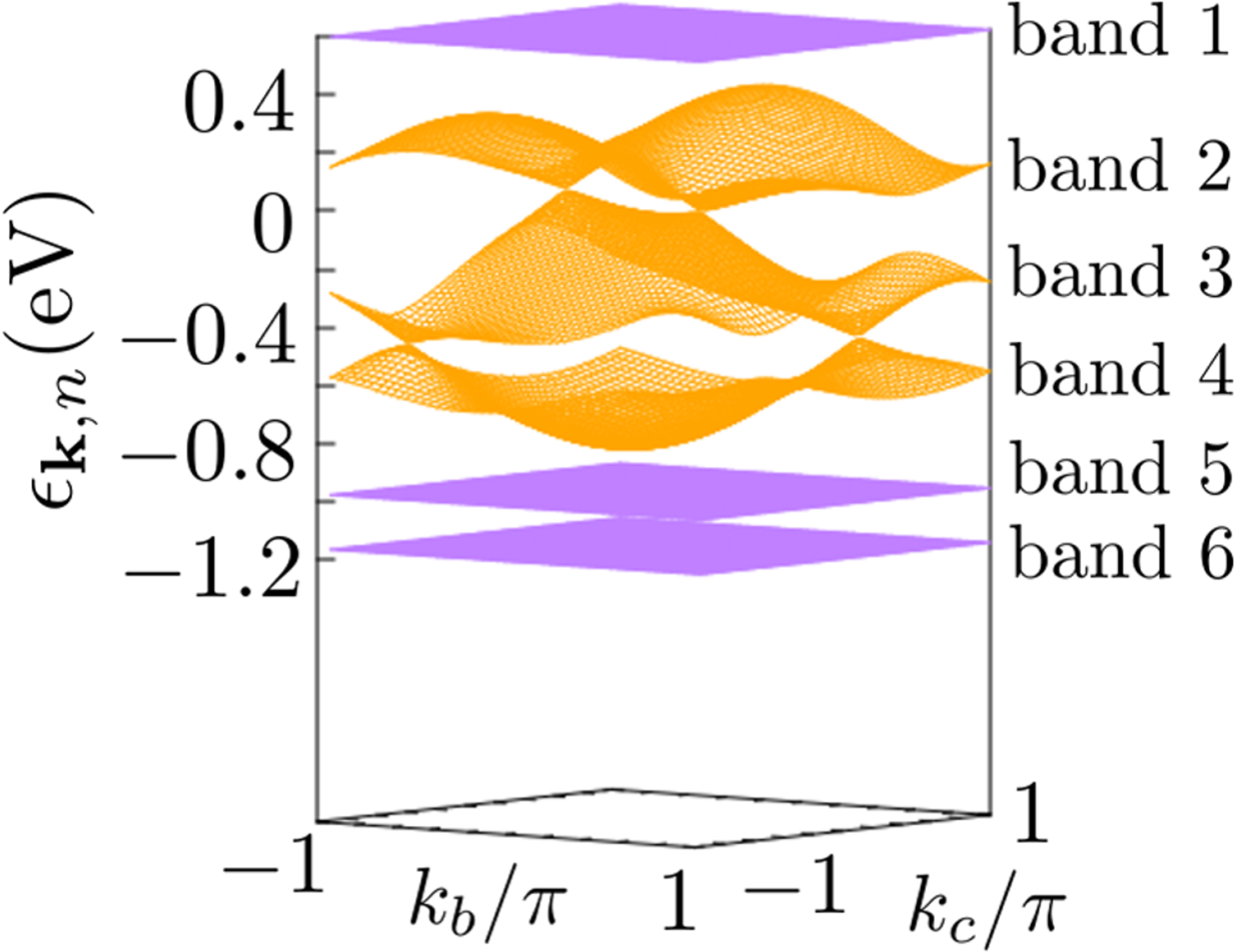}
\end{center}
\caption{(Color online) Energy dispersion obtained using six-orbital tight-binding model in $k_b$-$k_c$ plane, where $k_a$=$-\pi/2$. Bands 1, 5, and 6 are flat bands. 
} 
\label{dispersion_6}
\end{figure}

In Sect. 2, we developed a three-orbital tight-binding model based on Wannier fitting of the first-principles calculation. We also provide a detailed explanation on how the model was created and explain the process of creating three-orbital tight-binding models. Figure~\ref{DFT}(a) shows the energy band structure of [Ni(dmdt)$_2$], where the horizontal axis represents the connected symmetry points in the BZ. We performed Wannier fitting for the six energy bands near the Fermi energy and developed the six-orbital tight-binding model. Figure~\ref{wannier_6} visualizes the energy band structure obtained via Wannier fitting and first-principles calculations. The purple circles depict the Wannier fitting, whereas the red lines illustrate the first-principles calculation. The Wannier fitting reproduces the first-principles calculation. Using the hopping energies obtained via Wannier fitting, we created the six-orbital tight-binding model. We obtained the energy dispersion by diagonalizing the Hamiltonian of the six-orbital tight-binding model. Figure~\ref{dispersion_6} shows the energy dispersion of [Ni(dmdt)$_2$] in the $k_b$-$k_c$ plane, where $k_a$=$-\pi/2$. In Fig.~\ref{dispersion_6}, bands 1, 5, and 6 are flat bands and do not have the degenerate points between bands 2, 3, and 4. Thus, bands 2, 3, and 4 are separated into the three-orbital model. We then performed Wannier fitting for the three bands near the Fermi energy again and developed the three-orbital tight-binding model presented in Sect. 2. 

\end{document}